\documentclass[aps,prl,preprint,superscriptaddress]{revtex4}

\usepackage{graphicx}% Include figure files
\usepackage{dcolumn}% Align table columns on decimal point
\usepackage{bm}% bold math
\usepackage{color}

\newcommand{\dd}{\textrm{d}}
\newcommand{\dhh}{\textrm{d}h}
\newcommand{\dr}{\,\textrm{d}r}
\newcommand{\dx}{\,\textrm{d}x}
\newcommand{\dy}{\,\textrm{d}y}

\newcommand{\Ha}{\,\textit{Ha}}
\newcommand{\Pe}{\,\textit{Pe}}
\newcommand{\Ma}{\,\textit{Ma}}
\newcommand{\Ree}{\,\textit{Re}}
\newcommand{\Reem}{\,\textit{Re}_m}
\newcommand{\Ca}{\,\textit{Ca}}
\newcommand{\Sc}{\,\textit{Sc}}

\newcommand{\bB}{{\bm B}}

\newcommand{\bj}{{\bm j}}
\newcommand{\bn}{{\bm n}}
\newcommand{\bT}{{\bm T}}
\newcommand{\bu}{{\bm u}}
\newcommand{\bV}{{\bm V}}

\newcommand{\ii}{\textrm{i}}

\newcommand{\tB}{{\tilde B}}
\newcommand{\tc}{{\tilde c}}
\newcommand{\thh}{{\tilde h}}
\newcommand{\tu}{{\tilde u}}
\newcommand{\tphi}{{\tilde\phi}}

\newcommand{\cL}{\mathcal{L}}

\newcommand{\oL[1]}{\mathcal{L}\{{#1}\}}
\newcommand{\oLL[1]}{\mathcal{L}\{\mathcal{L}\{{#1}\}\}}

\begin{document}

\title{Electromagnetically driven flow in unsupported electrolyte
  layers: lubrication theory and linear stability of annular flow}

\author{Andrey Pototsky}
\email[]{apototskyy@swin.edu.au}
\affiliation{Swinburne University of Technology, Department of Mathematics, Hawthorn, 3122 Victoria, Australia}

%\thanks{}
\author{Sergey A.~Suslov}
\affiliation{Swinburne University of Technology, Department of Mathematics, Hawthorn, 3122 Victoria, Australia}
%\affiliation{}
%\author{}
%\thanks{}
%\affiliation{}
%\author{}
%\thanks{}
%\affiliation{}

\date{\today}
 
\begin{abstract}
We consider a thin horizontal layer of a non-magnetic electrolyte containing a bulk solution of salt and carrying an electric current. The layer is bounded by two deformable free surfaces loaded with an insoluble surfactant and is placed in a vertical magnetic field. The arising Lorentz force drives the electrolyte in the plane of the layer. We employ the long-wave approximation to derive general two-dimensional hydrodynamic equations describing symmetric pinching-type deformations of the free surfaces. These equations are used to study the azimuthal flow in an annular film spanning the gap between two coaxial cylindrical electrodes. In weakly deformed films, the base azimuthal flow and its linear stability with respect to azimuthally invariant perturbations are studied analytically. For relatively thick layers and weak magnetic fields, the leading mode with the smallest decay rate is found to correspond to a monotonic azimuthal velocity perturbation. The Marangoni effect leads to further stabilisation of the flow while perturbations of the solute concentration in the bulk of the fluid have no influence on the flow stability. In strongly deformed films in the diffusion-dominated regime, the azimuthal flow becomes linearly unstable with respect to an oscillatory mixed mode characterised by the combination of radial and azimuthal velocity perturbations when the voltage applied between electrodes exceeds the critical value.  
\end{abstract}


\maketitle

\section{Introduction}
\label{intro}
Electromagnetically driven flows in shallow layers and channels of
electrically conducting fluids in the presence of deformable
interfaces attracted much attention due to their importance in plasma
physics \cite{Fiflis16, Lunz19} and various microfluidic applications
including contactless manipulation of flow in magnetohydrodynamic
networks \citep{Bau03}, liquid channels embedded into carrier fluids
\citep{Dunne20}, droplet microfluidics \citep{Shang17} and
electromagnetic stirring \citep{Bau01, Qian05}.

The comprehensive theoretical description of magnetohydrodynamic (MHD)
flows of magnetic fluid films in arbitrary strong magnetic fields is
technically challenging as the hydrodynamic equations must be coupled
with Maxwell's equations in the presence of deformable moving
boundaries. However, in the case of non-magnetic electrically
conducting  fluids, the description can be greatly simplified. For
this class of fluids, which comprises electrolyte solutions and liquid
metals with weak magnetic properties, the additional stresses that
typically appear at the interfaces due to externally applied magnetic
fields can be neglected. For relatively weak magnetic fields of the
order of $10^{-2}-10^0$\,T,  which can be created using conventional
permanent magnets, the magnetic Reynolds number $\Reem=UL/\eta_m$
associated with the flow of fluid with the magnetic diffusivity
$\eta_m$ in the domain of a characteristic size $L$  with velocity
$U$, is typically small,$\Reem\ll1$. In this regime, the magnetic
field induced by the electric current flowing through the fluid can be
neglected compared to the external field \citep{Mueller13}. With
this simplification, the MHD equations have been successfully applied
to study flows of electrolitic solutions and non-magnetic liquid
metals such as mercury in various geometries. These include liquid
metal layers confined between two parallel insulating walls
\citep{Sommeria82} and in thin horizontal films \citep{Sommeria86} and
electrolyte solutions in annular channels \citep{Messadek02,
  Figueroa09, Perez16, Suslov17, McCloughan20}. 

Geometric parameters of the system such as the depth and the aspect
ration of the layer have been shown to have the major influence on the
flow characteristics. In shallow horizontal layers of electrolyte
solutions with a depth of several milimetres and a small
depth-to-width aspect ratio placed between two coaxial vertical
electrodes the flow was found to be essentially three-dimensional even
for relatively weak currents \citep{Figueroa09, Perez16,
  Suslov17, McCloughan20}. The quasi-two-dimensional
approximation, developed in \cite{Figueroa09, Perez16} by using
the depth-averaging method, could capture some of the main features of
the base azimuthal flow but was shown to be inadequate when describing
toroidal flows that lead to the formation of the experimentally
observable free-surface vortices \cite{Suslov17}.

As the depth of a horizontal layer and the aspect ratio of the system
are further decreased, the vertical component of the flow velocity is
impeded by the boundaries and the horizontal component of the flow
becomes dominant. The two-dimensional nature of the flow in very thin
liquid layers was used to study two-dimensional turbulence as
pioneered around four decades ago by \cite{Couder81, Couder84} in
experiments with soap films that were mechanically stirred by the
array of rods to produce turbulent flow. At around the same time
\cite{Sommeria86} studied effectively two-dimensional flows in thin
mercury films with a free upper surface and supported from below by an
array of conducting electrodes. Instead of a mechanical stirring, a
contactless electromagnetic Lorentz forcing was used to drive the flow
in the presence of highly non-uniform magnetic fields. Later, similar
contactless electromagnetic forcing was used to gain a deeper
understanding of the scaling properties of the velocity correlation
function in turbulent regimes \citep{Cardoso94, Marteau95,
  Williams97}. A comprehensive review of two-dimensional turbulence
can be found in \cite[e.g.]{Kellay02}. 

Historically, electromagnetic driving was extensively used in
supported liquid layers, but not in unsupported systems such as soap
films. This, perhaps, was due to the intrinsic instability of
soap-type films and the difficulty of controlling their curvature. In
fact, to the best of our knowledge, the first attempt to use Lorentz
force in unsupported free films was made almost 20 years after
\cite{Couder81, Couder84} pioneered the film turbulence
studies. It used a soap film containing chloride salt spanning a region between
two parallel conducting electrodes placed above an array of permanent
magnets \citep{Rivera00}. The main advantage of using an unsupported
film when studying two-dimensional turbulence is the elimination of energy
leakage at the no-slip bottom of the container. In a recent experimental
study \cite{Cruz16} investigated the flow dynamics in an
electromagnetically forced film in the case of a localised source of
electric current. A series of experiments with soap films spanning the
gap between two coaxial electrodes placed in an external magnetic
field is currently underway \cite{Cuevas23}.

The theoretical description and modelling of electromagnetically
driven flows in supported films with a free interface are now well
developed \citep{Morley95, Morley96, Morley97, Morley02, Gao02a,
  Gao02b, Miloshevsky10, Giannakis09a, Giannakis09b, Lunz19} and
continue to attract attention mainly due to applications in plasma
flows and tokamaks. In the absence of the Lorentz force, the pure
hydrodynamic description of the flow in unsupported liquid films 
was initiated in \cite{Prevost86, Sharma88} and later received a
huge boost because of its relevance to nonlinear film rupture
and two-dimensional turbulence problems \citep{Couder89, Chomaz90,
  Erneux93, Gharib98, Sharma95, VanDeFliert95, Wu95} as reviewed in
\citep{Kellay02, Oron97}. In unsupported thin viscous films, the hydrodynamic 
equations are simplified using two main assumptions. Firstly, the
long-wave approximation is applied by taking into account large-scale
flow patterns and film deformations the wavelength of which is much
larger than the average film thickness. Secondly, deformations of the
free interfaces are assumed to be mirror-symmetric with respect to
the centre plane of the layer, which corresponds to a varicose-type
pinching deformation mode. Under these assumptions, the effective
two-dimensional dynamic equations were derived at the leading order
of the lubrication approximation for curved soap films in the
presence of a surfactant \citep{Ida98, Miksis98} and for horizontally
stretched free films in the presence of solute and surfactant
\citep{Chomaz01}. So far, the application of the lubrication
approximation to describe MHD flows in unsupported free films of
electrolyte solutions in external magnetic fields has not been
reported.

Here we build upon earlier theoretical studies \citep{Ida98, Miksis98,
  Chomaz01} to derive the leading order dynamic equations for an
electromagnetically driven two-dimensional flow in a thin free
horizontal layer of electrolyte solution the free surfaces of which
are loaded with an insoluble surfactant. The flow is driven by the
Lorentz force generated by the electric current flowing through the
electrolyte in the presence of a homogeneous external magnetic field
normal to the layer. We show that at the leading order of the
lubrication approximation the product of the current density and the
local thickness of the layer is divergence-free reflecting the
condition of no accumulation of electric charge in the bulk. The
complete set of the derived dynamic equations is written in an
invariant vector form suitable for applications in arbitrary
geometries. It consists of the dynamic equation for the
two-dimensional flow field depending on the solute concentration in 
the bulk, surfactant concentration, symmetric film deformations and
the electric potential.  As an example, we apply the derived
equations to study the azimuthal flow and its linear stability in
annular free films spanning the gap between two coaxial conducting
electrodes.

The paper is organised as follows. In section~\ref{theory} we present
the derivation of the leading order equations in the lubrication
approximation using the systematic expansion technique suggested
earlier in \cite{Erneux93, Chomaz01}. The derived equations correctly
reflect the conservation of the total mass of the surfactant and
solute as well as the continuity of electric current under the
condition of no accumulation of the electric charge in the fluid. In 
section~\ref{stat} we rewrite the derived equations in polar
coordinates to study the flow in annular free films. In
section~\ref{stab} we study the linear stability of the annular
azimuthally invariant steady-state with respect to perturbations that
depend only on the radial coordinate. We present analytical results
for the linear stability of a flat film and vanishingly small flow
velocities first. Subsequently, we use the numerical continuation
method \cite{Krauskopf14, AUTO94} to study the stability of a strongly
deformed layer. The obtained theoretical and computational results are
summarised in section~\ref{concl}.

\section{Lubrication theory of electrically conducting free films in
  an external magnetic field}\label{theory}
\begin{figure} 
  \centerline{\includegraphics[width=\columnwidth]{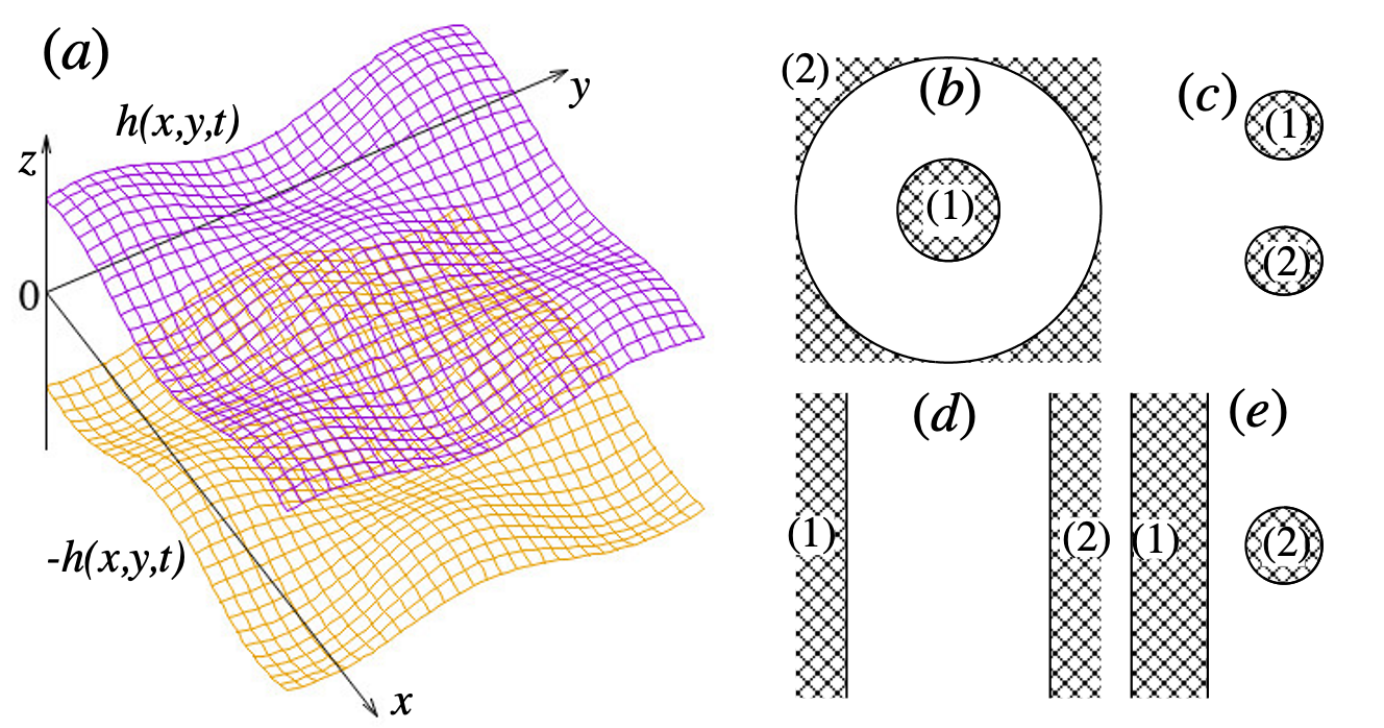}}
  \caption{($a$) Symmetric deformation mode in a horizontal free
    liquid film with two deformable surfaces located at $z=\pm
    h(x,y,t)$. The flow field ${\bm u}=(u,v,w)$ is mirror symmetric
    with respect to the centre-plane $z=0$. ($b$--$e$) The top view of
    the system: four possible topological configurations of a free
    film spanning space between two electrodes $(1,2)$. The surface of
    each electrode $\partial\Sigma$ represents a no-slip equipotential
    boundary impenetrable for surfactant and electrolyte solution.}
\label{fig0}
\end{figure}

Consider a horizontal free film of an electrolyte solution, which can
be created by supporting the weight of the film by the pressure
difference between the regions below and above the film. Following
\cite{Erneux93, Chomaz01} we exclude film bending and only consider
symmetric pinching-type surface deformation modes $z=\pm h(x,y,t)$
with each surface being a mirror image of the other at all times as
schematically shown in figure~\ref{fig0}(a). Here, $x$ and $y$ are the
coordinates in the horizontal plane, $z$ is a vertical coordinate and
$t$ is time. The local thickness of the film is $2h$ and the average
film thickness is
\begin{equation}
2\langle h\rangle=2S^{-1}\int_S h(x,y,t)\dx\dy\,,
\label{hmean}
\end{equation}
where $S$ denotes the area of the centre plane $z=0$. Each surface of
the film is loaded with an insoluble surfactant with local
concentration $c_s$. The addition of surfactants is particularly
important in soap films that contain fatty acid carboxylates, which
are typically found at the surface. The electrolyte solution is
composed of a solvent fluid (typically pure water) and dissociated
salt molecules with bulk concentration $c_b$. In what follows we
assume that salt is completely soluble and does not form a molecular
surface layer. Two electrodes are immersed in the fluid so that the electric
current can flow between them through the film when external voltage
is applied. One may consider at least four possible topological
configurations of a free film spanning space between two electrodes as
shown in figure \ref{fig0}($b$--$e$). The surface of the electrodes
$\partial\Sigma$ is assumed to be chemically inert and impenetrable to
the surfactant and the solute in the film. In addition, the flow field
${\bm u}$ vanishes at $\partial\Sigma$.
% such as sodium chloride \ce{NaCl} or potassium nitrate \ce{KNO3}. 

The electrical conductivity $\sigma$ of the electrolyte solution
generally depends on $c_b$. The solution density $\rho$ and dynamic
viscosity $\mu$ (kinematic viscosity $\nu=\mu/\rho$) are assumed to be
constant and independent of $c_b$. The externally applied magnetic
field $\bB=(0,0,B(x,y))$ is assumed to be significantly stronger than
that induced by the motion of the fluid. In what follows we neglect
gravity effects anticipating that hydrostatic pressure in a
sub-micrometre thin free film is negligible as compared to the Laplace
pressure. The motion of the incompressible fluid with
three-dimensional velocity $\bu=(u,v,w)$ is described by the
continuity and Navier-Stokes equations with the added Lorentz force
term \citep{Mueller13}
\begin{eqnarray}
  {\bm \nabla}\cdot\bu
  &=&0\,,\label{eq4a}\\
  \partial_t\bu+(\bu\cdot {\bm \nabla})\bu
  &=&-\frac{1}{\rho} {\bm \nabla} (p-\Pi)
      +\frac{\mu}{\rho}\nabla^2\bu+\frac{1}{\rho}\bj\times\bB\,,
      \label{eq1}
\end{eqnarray}
where $p$ is the pressure in the fluid and $\Pi=\Pi(h(x,y,t))$
represents the disjoining pressure due to intermolecular forces that
become important when the film thickness is approximately $100$\,nm or
less \cite{Israelachvili11, Overbeek60}. Because of the symmetry of
the pinching mode, the horizontal flow velocity components $(u,v)$ and
the vertical velocity component $w$ should be even and odd function of
$z$, respectively.
%More specifically, we use the combination of the attracting
%long-range van der Waals forces and repulsive short-range
%electrostatic forces induced by an electric double-layer
%\citep{Overbeek60} 
%\begin{eqnarray}
%\label{eq1a}
%\psi=\frac{A}{6\pi (2h)^3}-\kappa Ee^{-2\kappa h},
%\end{eqnarray}
%where $A$ is the Hamaker constant, $E$ is the excess energy of the
%double-layer forces per unit area and $\kappa$ is the reciprocal of
%the Debye length associated with the double-layer. Note that $\psi$
%is the function of the local film thickness only and is independent
%of $z$. 

The current density $\bj=(j_x,j_y,j_z)$ is related to the electric
potential $\phi$ via Ohm's law 
\begin{equation}
  \bj=\sigma(c_b) (-{\bm \nabla}\phi+\bu\times\bB)\,.\label{eq2}
\end{equation}
For electrolyte solutions, we assume linear dependence between
conductivity $\sigma$ and $c_b$
\begin{equation}
  \sigma(c_b)=Kc_b\,,\label{eq_sigma}
\end{equation}
where $K$ is an empirical constant specific to a particular salt and
solvent. Under the condition of no accumulation of electric charge in
the bulk, the electric potential $\phi$ is found from 
\begin{equation}
  {\bm \nabla}\cdot\bj=0\,,\label{eq3}
\end{equation}
which must be solved instantaneously for any given velocity field
$\bu$. Additionally, the current continuity condition \ref{eq3} must
be supplemented with the boundary conditions for $\phi$ that
correspond to the equipotential surfaces $\partial\Sigma$ of the
conducting electrodes.
% Note that \ref{eq3} can be obtained from \ref{eq2} under the
% assumption of no local accumulation of charges,
% i.e.~${\bm \nabla}\cdot\bj=0$.

Note that for a magnetic field $\bB=(0,0,B)$ orthogonal to the layer
$\bj\times\bB=B(j_y,-j_x,0)$,
$\bj=\sigma(c_b)(-\partial_x\phi+Bv,-\partial_y\phi-Bu,-\partial_z\phi)$
and $\bu\times\bB=B(v,-u,0)$. The normal component of the electric
current vanishes at the film surfaces. Thus, at t $z=h$ we require
\begin{equation}
  j_n=\bj\cdot\bn=0\,,\label{eq4b}
\end{equation}
where $\bn=(-\partial_xh,-\partial_yh,1)
/\sqrt{1+(\partial_x h)^2+(\partial_y h)^2}$ is the unit normal vector
to the upper film surface directed away from the fluid.

At $z=h$, the kinematic boundary condition applies
\begin{equation}
  \partial_t h+(\bu\cdot {\bm \nabla}_\parallel)h=w\,,\label{eq5}
\end{equation}
where ${\bm \nabla}_\parallel=(\partial_x,\partial_y)$ is the horizontal
gradient. Note that \ref{eq5} can also be written in the equivalent
form as $\partial_t h=\sqrt{1+({\bm \nabla}_\parallel h)^2}(\bn\cdot\bu)$.

To describe the dynamics of surfactant and the concentration of salt
in the bulk of the fluid, we follow \cite{Jensen93}. The bulk
concentration $c_b$ is described by the advection-diffusion equation 
\begin{equation}
  \partial_tc_b+{\bm \nabla}\cdot(\bu c_b-d_b{\bm \nabla} c_b)=0\,,\label{eq_cb}
\end{equation}
where $d_b$ is the bulk diffusion coefficient, $\bu c_b$ is the
advective flux and $-d_b{\bm \nabla} c_b$ is the diffusive flux.
%Using the continuity condition (\ref{eq4a}), 

The advection-diffusion equation for the surfactant concentration at
the upper film surface $z=h(x,y,t)$ is given by 
\begin{equation}
  \partial_t c_s+{\bm \nabla}_s\cdot(c_s\bu)=d_s{\bm \nabla}_s^2c_s\,,\label{eq4}
\end{equation}
where $d_s$ is the surface diffusivity of the surfactant and
${\bm \nabla}_s={\bm \nabla}-\bn(\bn\cdot {\bm \nabla})$ is the surface gradient.

At $z=h$ the diffusive flux normal to the film surface must vanish 
\begin{equation}
  \bn\cdot {\bm \nabla} c_b=0\,.\label{eqcb1}
\end{equation}
%Additionally, at the surface of the electrodes emerged in the fluid,
%both, the flow velocity ${\bm u}$ and the normal component of the
%diffusive flux $-d_b{\bm \nabla} c_b$ must vanish. 
%Integrating (\ref{eq_cb}) over $z$ between $z=0$ and $z=h$ and then
%over the entire area of the centre surface $S$ between the electrodes
%and making use of the divergence theorem, we obtain  
%\begin{eqnarray}
%\label{eq_vol}
%\int_{S}\dx\dy\int_0^{h(x,y,t)}\dz\partial_t c_d+
%\int_{\Sigma} ({\bm u}c_b-d_b{\bm \nabla} c_b)\cdot {\bm n} d\Sigma=0 
%\end{eqnarray}
%where $\Sigma$ represents the time-dependent surface area of the film.
%Finally, using the boundary condition (\ref{eqcb1}), the kinematic
%condition $\partial_t h =\sqrt{1+({\bm \nabla}_\parallel h)^2}({\bm
%n}\cdot {\bm u})$, the identity $\partial_t \int_0^{h(t)} f(t,z)dz =
%\int_0^{h(t)} \partial_t f(t,z)dz + f(t,h(t))\partial_t h(t)$, and
%substituting $d\Sigma=\sqrt{1+({\bm \nabla}_\parallel
%h)^2}\dx\dy$, we arrive at the conservation of the total
%volume of the dissolved salt 
It can be shown that \ref{eqcb1} supplemented with the condition
that the flow velocity $\bu$ and the normal diffusive fluxes of
surfactant and solute vanish at the surface of the electrodes leads to
the conservation of the total mass of the solute and the surfactant. 
%\begin{eqnarray}
%\label{eq_vol1}
%2\partial_t \left(\int_{S}\dx\dy
%\int_0^{h(x,y,t)}c_d\dz  \right)=0,~~2\partial_t
%\left(\int_{S}\dx\dy c_s \right)=0. 
%\end{eqnarray}
%Similarly, integrating (\ref{eq4}) over the entire surface area of
%the film $\Sigma$, we obtain the conservation of the total mass of
%insoluble surfactant  $2\partial_t \left(\int_{S}\dx\dy
%c_s \right)=0$ as long as the flow velocity ${\bm u}$ and the normal
%component of the surface diffusive flux $-d_s{\bm \nabla}_s c_s$
%vanish at the boundary. 
%Additionally, we assume that there is no local accumulation of
%charges in the bulk of the fluid, which yields an additional
%condition for the current 
%\begin{eqnarray}
%\label{eq4c}
%{\bm \nabla }\cdot {\bm j}=0.
   %  \end{eqnarray}

Next, we consider the balance of the normal and tangential forces at
the upper surface $z=h(x,y,t)$
\begin{equation}
  (p+\Gamma\kappa)\bn={\bm \nabla}_s\Gamma+\bT\cdot\bn\,,\label{eq6}
\end{equation}
where $\bT=\mu[{\bm \nabla}\otimes\bu+({\bm \nabla}\otimes\bu)^T]$ is the
viscous stress tensor,  $\kappa(x,y,t)$ is the local mean curvature of
the surface defined as $\kappa=-{\bm \nabla}\cdot\bn$, $\Gamma(x,y,t)$ is
the local surface tension,  $\otimes$ is the tensor product and the
superscript $T$ denotes transposed quantities. In what follows we
assume that the liquid is non-magnetic and the applied magnetic field
is relatively weak so that the Maxwell component in the stress tensor
can be completely neglected.
  
The gradient of the surface tension along the interface
$\partial_s\Gamma$ is induced by the distribution of the surfactant
according to the soluto-Marangoni effect
\begin{equation}
  \Gamma(x,y,t)=\gamma-\Gamma_Mc_s(x,y,t)\,,\label{eq7}
\end{equation}
where $\gamma$ is the reference surface tension in the absence of a
surfactant and $\Gamma_M=-d\Gamma/dc_s>0$ is assumed to be
constant. The balance of forces at the lower surface $z=-h(x,y,t)$ is
automatically achieved for symmetric deformation modes. 

The horizontal length scale $L$ of the flow in submicrometre thin
films is several orders of magnitude larger than the average film
thickness $2\langle h\rangle$. The long-wave approximation theory of
one-dimensional free films in the absence of a surfactant and an
electric current has been developed some thirty years ago
\citep{Erneux93} using systematic expansion of the Navier-Stokes
equations for small lubrication parameter $\epsilon=\langle h\rangle
/L \ll 1$. Subsequently, the theory was generalised to describe
two-dimensional flat and curved free films loaded with soluble and
insoluble surfactant agents \citep{Ida98, Miksis98, Chomaz01}. Here we
extend earlier results to derive the leading-order equations for
long-wave symmetric deformations of a free electrically conducting
film placed in an external magnetic field. 

We scale horizontal coordinates $(x,y)$ with $L$ and the vertical
coordinate $z$ and the local interface deflection $h(x,y,t)$ with
$\langle h\rangle=\epsilon L$. The horizontal fluid velocity $(u,v)$
is scaled with some reference velocity $U=O(1)$, the vertical velocity
$w$ with $\epsilon U=O(\epsilon)$, time with $L/U=O(1)$ and the
pressure and the disjoining pressure with $\rho U^2$. The magnetic
field $B(x,y)$ is non-dimensionalised using some reference value $\tB$
while the scaling for the electric potential is $U\tilde{B}L$. The bulk
salt and the surface surfactant concentrations are scaled using
arbitrary reference concentrations $c_b^{(0)}$ and $c_s^{(0)}$,
respectively, so that the conductivity $\sigma(c_b)$ is scaled with
$Kc_b^{(0)}$. The dimensionless bulk equations~\ref{eq4a},
\ref{eq1}, \ref{eq3} and \ref{eq_cb} then become 
\begin{eqnarray}
  \partial_xu+\partial_yv+\partial_z w
  &=&0\,,\label{eq8d}\\
  \partial_tu+(u\partial_x+v\partial_y+w\partial_z)u
  &=&-\partial_x[p-\Pi]
      +\Ree^{-1}\left(\partial_x^2+\partial_y^2+\epsilon^{-2}\partial_z^2\right)u
      \nonumber\\
  &&-\Ha^2\Ree^{-1}c_bB(Bu+\partial_y\phi)\,,\label{eq8a}\\
  \partial_tv+(u\partial_x+v\partial_y+w\partial_z)v
  &=&-\partial_y[p-\Pi]
      +\Ree^{-1}\left(\partial_x^2+\partial_y^2+\epsilon^{-2}\partial_z^2\right)v 
      \nonumber\\
  &&-\Ha^2\Ree^{-1}c_bB(Bv-\partial_x\phi)\,,\label{eq8b}\\
  \partial_t w+(u\partial_x+v\partial_y+w\partial_z)w
  &=&-\epsilon^{-2}\partial_z
      p+\Ree^{-1}\left(\partial_x^2+\partial_y^2+\epsilon^{-2}\partial_z^2\right)w\,, 
      \label{eq8c}\\
      \partial_x[c_b(Bv-\partial_x\phi)]
  &-&\partial_y[c_b(Bu+\partial_y\phi)]
      -\epsilon^{-2}\partial_z[c_b\partial_z\phi]=0\,,\label{eq8e}\\
    \partial_tc_b+u\partial_xc_b+v\partial_yc_b+w\partial_zc_b
  &=&\Sc^{-1}\Ree^{-1}
      \left(\partial_x^2+\partial_y^2+\epsilon^{-2}\partial_z^2\right)c_b\,,
      \label{eq8f}
\end{eqnarray}
where we used the same notations for the dimensionless quantities and
introduced the Reynolds ($\Ree$), Hartmann ($\Ha$), Peclet ($\Pe$), 
and Schmidt ($\Sc$) numbers
\begin{equation}
  \Ree=\frac{UL\rho}{\mu}\,,\quad \Ha^2
  =\frac{\tB^2L^2Kc_b^{(0)}}{\mu}\,,\quad
  \Pe=\frac{LU}{d_s}\,,\quad
   \Sc=\frac{\mu}{\rho d_b}\,.\label{eq9}
\end{equation}
From equation~\ref{eq6}, the dimensionless normal and the tangential
balances of stresses at $z=h$ are given by 
\begin{eqnarray}
  p+ (\Ca\Ree)^{-1}\left(\partial_x^2h+\partial_y^2h\right)
  &=&2\Ree^{-1}\left(\partial_zw-\partial_xh\,\partial_zu
      -\partial_yh\,\partial_zv\right)+O(\epsilon^2)\,,\label{stress1}\\
  -\Ma\Pe^{-1}\partial_xc_s
  &=&\epsilon^{-2} \partial_zu+\left[-2\partial_xh\,\partial_xu
      -\partial_yh(\partial_yu+\partial_xv)+\partial_xw\right.\nonumber\\
  &&-\left.(\partial_xh)^2\partial_zu-\partial_xh\,\partial_yh\,\partial_zv
      +2\partial_xh\,\partial_zw\right]+O(\epsilon^2)\,,\label{stress2}\\
  -\Ma\Pe^{-1}\partial_yc_s
  &=&\epsilon^{-2}\partial_zv+\left[-2\partial_yh\,\partial_yu
      -\partial_xh(\partial_xv+\partial_yu)+\partial_y\,w\right.\nonumber\\
  &&-\left.(\partial_yh)^2\partial_zv-\partial_xh\,\partial_yh\,\partial_zv
      +2\partial_yh\,\partial_zw\right]+O(\epsilon^2)\,,
  \label{stress3}
\end{eqnarray}
with the capillary ($\Ca$) and Marangoni ($\Ma$) numbers defined as
\begin{equation}
  \Ca=\frac{\mu UL}{\langle h\rangle\gamma}\,,\quad
  \Ma=\frac{c_s^{(0)}\Gamma_ML^2}{\langle h\rangle\mu d_s}\,.
  \label{Ca_Mar}
\end{equation}
The scaled boundary condition for the electric current~\ref{eq4b},
the kinematic condition~\ref{eq5}, the surfactant
equation~\ref{eq4} and \ref{eqcb1} multiply by
$\sqrt{1+(\partial_xh)^2+(\partial_y h)^2}$ are given by 
\begin{eqnarray}
  0
  &=&(\partial_x\phi-Bv)\partial_x h
      +(\partial_y\phi-Bu) \partial_y h
      +\epsilon^{-2}\partial_z\phi\,,\label{bcs}\\
  \partial_th
  &=&-u\partial_xh-v\partial_yh+w\,,\label{eq8h}\\
  \partial_tc_s
  &=&-\partial_x[uc_s]-\partial_y[vc_s]
      +\Pe^{-1}\left(\partial_x^2+\partial_y^2\right)c_s
      +O\left(\epsilon^2\right)\,,\label{eq8g}\\
  0
  &=&\partial_xh\,\partial_xc_b+\partial_yh\,\partial_yc_b
       -\epsilon^{-2}\partial_zc_b\,.\label{bcs1}
\end{eqnarray}

Crucial for further analysis is to determine the order of magnitude of 
all dimensionless parameters appropriate for the physical regime of
interest. Following \cite{Chomaz01} we assume that for free liquid
films the inertial effects play an essential role implying that
$\Ree=O(1)$. The leading contribution to pressure in the fluid is
anticipated to come from the Laplace pressure, which implies that
$\Ca=O(1)$. Note that, for example, for slipper bearing flows and
liquid films on a solid substrate, the capillary number scales
as $\Ca=(U\mu/\gamma)\epsilon^{-3}=O(1)$ \citep{Oron97}. We assume
that the Marangoni effect is weak so that at the leading order the
film surfaces can be considered stress-free \citep{Chomaz01}. This can
be achieved by setting $\Ma=O(1)$. An additional assumption must be
made regarding the strength of the magnetic field and the induced
electric current. Here we consider weakly conducting electrolytes in
weak magnetic fields and assume that the Lorentz force is of the same
order of magnitude as the viscous force in the absence of vertical
shear, i.e.~${\bm \nabla}_\parallel^2u\sim\Ha^2{\bm \nabla}_\parallel\phi$. This
implies that $\Ha=O(1)$. 

All fields in equations~(\ref{eq8d})--(\ref{bcs1}) are then expanded
into a series in powers of $\epsilon^2$,
e.g.~$u=u_0+\epsilon^2u_1+\ldots$ with $u_i=O(1)$, and the leading
zero-order equations are derived by following the procedure outlined
in \cite{Erneux93} and \cite{Chomaz01}. At the zeroth order, the
equations for $h_0$, $u_0$, $v_0$, $w_0$, $p_0$, and $(c_{s})_0$ are
identical to those derived in \cite{Chomaz01} as the Lorentz force
enters the equations only at the next order. Therefore, $u_0$, $v_0$
and $p_0$ are independent of $z$ and $w_0=-(\partial_x u_0+\partial_y
v_0)z$.

At the leading order, from equations~\ref{eq8e} and \ref{eq8f} we
obtain for the electric potential $\phi_0$ and the bulk concentration
$(c_{b})_0$  
\begin{equation}
  \partial_z[(c_b)_0\,\partial_z\phi_0]=0\,,\quad
  \partial_z^2(c_{b})_0=0\,.\label{eq9a}
\end{equation}
Since $\partial_z\phi_0=\partial_z(c_{b})_0=0$ at $z=h_0$, from
(\ref{bcs}) and (\ref{bcs1}) we conclude that both $(c_{b})_0$  and
$\phi_0$ are independent of $z$.

The surfactant concentration $(c_{s})_0$ satisfies the two-dimensional
advection-diffusion equation 
\begin{equation}
  \partial_t(c_s)_0{\bm \nabla}_\parallel\cdot[(c_s)_0\bu_0]
  =\Pe^{-1}{\bm \nabla}_\parallel^2(c_s)_0\,.\label{eq9aa}
\end{equation}
where $\bu_0=(u_0,v_0)$ 
is the leading order horizontal velocity. The kinematic
equation~\ref{eq8h} together with
$w_0(z=h_0)=-({\bm \nabla}_\parallel\cdot\bu_0)h_0$ yield the evolution
equation for the local film deformation
\begin{equation}
  \partial_th_0+{\bm \nabla}_\parallel\cdot[h_0u_0]=0\,.\label{eq9b}
\end{equation}

At the next order, the Navier-Stokes equations for the horizontal flow
contain the Lorentz force terms 
\begin{eqnarray}
  \partial_tu_0+(u_0\partial_x+v_0\partial_y)u_0
  &=&-\partial_x[p_0-\Pi_0]
      +\Ree^{-1}\left(\partial_x^2+\partial_y^2\right)u_0
      -\Ha^2\Ree^{-1}(c_b)_0B_0(u_0+\partial_y\phi_0)\nonumber\\
  &&+\Ree^{-1}\partial_z^2u_2\,,\\
  \partial_t v_0+(u_0\partial_x+v_0\partial_y)v_0
  &=&-\partial_y[p_0-\Pi_0]
      +\Ree^{-1}\left(\partial_x^2+\partial_y^2\right)v_0
      -\Ha^2\Ree^{-1}(c_b)_0B_0(v_0-\partial_x\phi_0)\nonumber\\
  &&+\Ree^{-1}\partial_z^2v_2\,,\label{eq10}
\end{eqnarray}
where the electric potential $\phi_2$ satisfies the equation
\begin{equation}
  \partial_x[(c_{b})_0(B_0v_0-\partial_x\phi_0)]
  -\partial_y[(c_b)_0(B_0u_0+\partial_y\phi_0)]
  +(c_{b})_0\partial_z^2\phi_2=0\,.\label{eq11}
\end{equation}
The boundary condition for $\phi_2$ at the film surface $z=h_0$
obtained from equation~\ref{bcs} is  
\begin{equation}
  \partial_z\phi_2=(B_0u_0+\partial_y\phi_0)\partial_yh_0
  -(B_0v_0-\partial_x\phi_0)\partial_xh_0\,.\label{eq11a}
\end{equation}
Equation~\ref{eq11} shows that $\phi_2$ is a quadratic function of
$z$ 
\begin{equation}
  \phi_2=A(x,y)z^2+B(x,y)z+C(x,y)\,,\label{eq11b}
\end{equation}
Since for symmetric deformations the potential $\phi$ must be an even
function of $z$, we set $B(x,y)=0$ and determine $A(x,y)$ from the
boundary condition~\ref{eq11a} to obtain
\begin{equation}
  \phi_2=-\frac{[(B_0v_0-\partial_x\phi_0)\partial_xh_0
    -(B_0u_0+\partial_y\phi_0)]z^2\partial_yh_0}{2h_0}
  +C(x,y)\,.\label{eq11c}
\end{equation}
Finally, substituting \ref{eq11c} into \ref{eq11} we obtain
\begin{eqnarray}
  \partial_x[(c_{b})_0(B_0v_0-\partial_x\phi_0)]
  &-&\partial_y[(c_{b})_0(B_0u_0+\partial_y\phi_0)]\nonumber\\
  &+&\frac{(c_{b})_0[(B_0v_0-\partial_x\phi_0)\partial_xh_0
      -(B_0u_0+\partial_y\phi_0)\partial_yh_0]}{h_0}=0\,.
      \label{eq11d}
\end{eqnarray}
Multiplying equation~\ref{eq11d} by $h_0$ and introducing the
current $\bj_0=(c_b)_0(\bu_0\times\bB_0-{\bm \nabla}_\parallel\phi_0)$ we
rewrite equation~\ref{eq11d} in an invariant vector form 
\begin{equation}
  {\bm \nabla}_\parallel\cdot(h_0\bj_0)=0\,.\label{eq11e}
\end{equation}
Equation~\ref{eq11e} represents the continuity equation for the
electric current per unit length of the cross-section of the film.
%Next, we note that the continuity condition (\ref{bcs1}) at the
%leading order exactly coincides with (\ref{eq9a}). 
%In the next order, we obtain from (\ref{bcs1})
%\begin{eqnarray}
%\label{cont_eq1}
%0&=&\partial_x(-\partial_x \phi_0 +v_0)+\partial_y(-\partial_y\phi_0
%-u_0)-\partial_z^2 \phi_2. 
%\end{eqnarray}
%Using the expression for $\phi_2$ from (\ref{eq11c}), the equation
%(\ref{cont_eq1}) becomes 
%\begin{eqnarray}
%\label{cont_eq2}
%0&=&\partial_x(-\partial_x \phi_0 +v_0)+\partial_y(-\partial_y \phi_0
%-u_0)+\frac{[\partial_x \eta_0 (-\partial_x \phi_0 +v_0)+\partial_y
%\eta_0 (-\partial_y \phi_0 -u_0) ]}{1+\eta_0}\nonumber\\ 
%\end{eqnarray}
%Multiplying equation (\ref{cont_eq2}) with $1+\eta_0$, we finally
%obtain the continuity equation for the current ${\bm
%j}_0=(-\partial_x \phi_0 +v_0,-\partial_y \phi_0 -u_0)$ 
%\begin{eqnarray}  
%\label{cont_eq3}
%0&=&{\bm \nabla}_\parallel \cdot [(1+\eta_0){\bm j}_0].
%\end{eqnarray}
%It is interesting to observe that with the condition (\ref{cont_eq3})
%the equation for the electric potential (\ref{eq11e}) is identically
%fulfilled. This shows that in the lubrication 

Next, we eliminate $u_2$ and $v_2$ from equations~\ref{eq10} by
taking into account the boundary conditions for the tangential and
normal components of the stress tensor at $z=h_0$. Because the
electromagnetic component of the viscous stress tensor is neglected
here, the result of the elimination procedure is identical to that of
\cite{Chomaz01}. Consequently,
%introducing the vector potential ${\bm \Phi}_0=(0,0,\phi_0)$,
%and noticing that $(\partial_y\phi_0 ,-\partial_x\phi_0,0)
%=\nabla_\parallel\times{\bm \Phi}_0$,
%\begin{eqnarray}
%\label{eq11f}
%(\partial_y \phi_0 ,-\partial_x \phi_0,0)={\nabla}_\parallel
%\times {\bm \Phi},
%\end{eqnarray}
we arrive at the leading-order dynamic equation for $\bu_0$ including
the Lorentz force
\begin{eqnarray}
  \partial_t\bu_0
  &+&(\bu_0\cdot{\bm \nabla}_\parallel)\bu_0
      =\frac{\dd\Pi_0}{\dhh_0}{\bm \nabla}_\parallel h_0
      +(\Ca\Ree)^{-1}{\bm \nabla}_\parallel\nabla_\parallel^2h_0
  +3\Ree^{-1}{\bm \nabla}_\parallel({\bm \nabla}_\parallel\cdot\bu_0)\nonumber\\
  &+&\Ree^{-1}\nabla_\parallel^2\bu
      -\frac{(\Ree\Pe)^{-1}\Ma}{h_0}{\bm \nabla}_\parallel(c_{s})_0
      +\frac{\Ree^{-1}}{h_0}\bV
      +\Ha^2\Ree^{-1}\bj_0\times\bB_0\,,\label{eq12}
\end{eqnarray}
where we replaced the vector
$(c_b)_0B_0[(-\partial_y\phi_0,\partial_x\phi_0)-B_0\bu_0]$ with
$\bj_0\times\bB_0$ and introduced an additional flow field
\begin{equation}
  \bV=2({\bm \nabla}_\parallel h_0\cdot{\bm \nabla}_\parallel)\bu_0
  +{\bm \nabla}_\parallel h_0\times({\bm \nabla}_\parallel\times\bu_0)
  +2{\bm \nabla}_\parallel h_0({\bm \nabla}_\parallel\cdot\bu_0)\label{eq13}
\end{equation}
that can be associated with the so-called extensional Trouton viscosity.

To close the system of leading order dynamic equations we derive the
equation for the bulk concentration $(c_b)_0$. At the leading order,
from \ref{eq9a} we see that $(c_b)_0$ is independent of $z$. At the
next order, from \ref{eq8f} we obtain 
\begin{equation}
  \partial_t(c_b)_0+u_0\partial_x(c_b)_0+v_0\partial_y(c_b)_0
  =\Sc^{-1}\Ree^{-1}(\partial_x^2+\partial_y^2)(c_b)_0
  +\Sc^{-1}\Ree^{-1}\partial_z^2(c_b)_2\,.\label{cb1}
\end{equation}
The boundary condition at $z=h_0$ following from \ref{bcs1} reads
\begin{equation}
  \partial_z(c_b)_2=\partial_xh_0\partial_x(c_b)_0
  +\partial_yh_0\partial_y(c_b)_0\,.\label{cb2}
\end{equation}
For a symmetric mode and according to \ref{cb1} the field $(c_b)_2$
is a quadratic function of $z$:
$(c_b)_2=a(x,y,t)z^2+c(x,y,t)$. Applying \ref{cb2} we write 
\begin{equation}
  (c_b)_2=\frac{\partial_xh_0\,\partial_x(c_b)_0
    +\partial_yh_0\,\partial_y(c_b)_0}{2h_0}z^2
  +c(x,y,t)\,.\label{cb3}
\end{equation}
Substituting \ref{cb3} into \ref{cb1} we obtain
\begin{eqnarray}
  \partial_t(c_b)_0
  &+&u_0\partial_x(c_b)_0+v_0\partial_y(c_b)_0
      =\Sc^{-1}\Ree^{-1}(\partial_x^2+\partial_y^2)(c_b)_0\nonumber\\
  &+&\Sc^{-1}\Ree^{-1}\frac{\partial_xh_0\,\partial_x(c_b)_0
      +\partial_yh_0\,\partial_y(c_b)_0}{h_0}\,.\label{cb4}
\end{eqnarray}
Multiplying \ref{cb4} by $h_0$ and using the kinematic condition
\ref{eq9b} we arrive at  
\begin{equation}
  \partial_t [h_0(c_b)_0]
  +{\bm \nabla}_\parallel\cdot[h_0\bu_0(c_b)_0]
  =\Sc^{-1}\Ree^{-1}{\bm \nabla}_\parallel\cdot[h_0{\bm \nabla}_\parallel
  (c_b)_0]\,.\label{cb5}
\end{equation}
Equation~\ref{cb5} is identical to the transport equation for the
solute concentration obtained at the leading order of the lubrication
approximation after averaging over the film cross-section that was
derived in \cite{Jensen93}. It generalises the leading order equation
for the bulk concentration derived in \cite{Chomaz01} to the case of
large film deformations. 
%Indeed, 
%integrating (\ref{cb5}) over the entire area $S$ of the centre
%surface of the film, we see that the total solute mass  $\int_S
%\dx\dy (1+\eta_0)(c_b)_0$ is conserved as long as the
%flow velocity ${\bm u}_0$ and the normal component of the diffusive
%flux ${\bm \nabla}_\parallel (c_b)_0$ vanish at the boundary of
%$S$. In practice, the film is spanned between two electrodes with a
%no-slip boundary impenetrable to surfactants and solutes, which
%automatically ensures the conservation of the total solute mass. 
%Equations~(\ref{eq12}) and (\ref{eq13}) are complemented with
%equations~(\ref{eq9a}), (\ref{eq9b}), (\ref{eq11}).  

We summarise our results by writing the complete set of dimensional
governing equations using physical variables and parameters 
\begin{eqnarray}
  \rho(\partial_t\bu+(\bu\cdot{\bm \nabla}_\parallel)\bu)
  &=&g(h){\bm \nabla}_\parallel h
      +\gamma{\bm \nabla}_\parallel{\bm \nabla}_\parallel^2h
      +\mu{\bm \nabla}_\parallel^2\bu+3\mu{\bm \nabla}_\parallel
      ({\bm \nabla}_\parallel\cdot\bu)\nonumber\\
  &&-\frac{\Gamma_M}{h}{\bm \nabla}_\parallel c_s+\frac{\mu\bV}{h}
     +\bj\times\bB\,,\nonumber\\
  \bV
  &=&2({\bm \nabla}_\parallel h\cdot{\bm \nabla}_\parallel)\bu
      +{\bm \nabla}_\parallel h\times ({\bm \nabla}_\parallel\times\bu)
      +2{\bm \nabla}_\parallel h({\bm \nabla}_\parallel\cdot\bu)\,,\nonumber\\
  {\bm \nabla}_\parallel\cdot(h\bj)&=&0\,,\nonumber\\
  \bj&=&Kc_b[\bu\times\bB-{\bm \nabla}_\parallel\phi]\,,\nonumber\\
%{\bm \nabla}_\parallel^2\phi&=& ({\bm \nabla}_\parallel\times\bu)\cdot\bB
%+\frac{{\bm \nabla}_\parallel h\cdot(-{\bm \nabla}_\parallel\phi
%+\bu\times\bB)}{h}\,,\nonumber\\ 
%\bj&=&\sigma\left(-{\bm \nabla}\phi+\bu\times\bB\right)\,,\nonumber\\ 
  \partial_t h&=&-{\bm \nabla}_\parallel\cdot(h\bu)\,,\nonumber\\
  \partial_t(h c_b)
  &=&-{\bm \nabla}_\parallel\cdot(hc_b\bu)
      +d_b{\bm \nabla}_\parallel\cdot(h{\bm \nabla}_\parallel c_b)\,,\nonumber\\
  \partial_tc_s
  &=&-{\bm \nabla}_\parallel\cdot(c_s\bu)+d_s{\bm \nabla}_\parallel^2c_s\,,
      \label{phys_eq1}
\end{eqnarray}
where we introduced function $g(h)=\dd\Pi(h)/\dhh$
and dropped subscript $0$. Note that equations~\ref{phys_eq1}
are written in a compact vector form, which is invariant with respect
to the choice of a coordinate system. This is especially important in
applications with non-rectangular geometry as exemplified in the next
section.

\section{Azimuthal flow in an annular free film between two coaxial
  cylinders}\label{stat}
Electromagnetically driven flows of electrolytes in annular bounded
by cylindrical vertical electrodes and a solid bottom have been
extensively studied experimentally and theoretically
\citep[.e.g.]{Perez16, Suslov17, McCloughan20}. It was found that the
steady azimuthal flow may become unstable giving rise to free-surface
vortices developing close to the outer cylindrical wall. The steady
flow field has a three-dimensional toroidal structure while the
deformation of the upper free surface is negligible. 

However, in thin liquid layers the film deformation can no longer be
neglected as has demonstrated in \citep{Wu95} using a mechanically
driven flow in an unsupported soap film spanning the gap between two
thin coaxial discs. If the outer disc is fixed and the inner one is
rotated, the fluid is set in motion in an azimuthal direction similar
to a Couette cell flow. Centrifugal forces push the liquid towards the
outer disc making the film thinner near the inner disc. Rather
unexpectedly, the flow was found to be laminar and the onset of
turbulence was not observed even at the linear rotation speed of up to
3\,ms$^{-1}$.

\begin{figure} 
  \centerline{\includegraphics[width=\columnwidth]{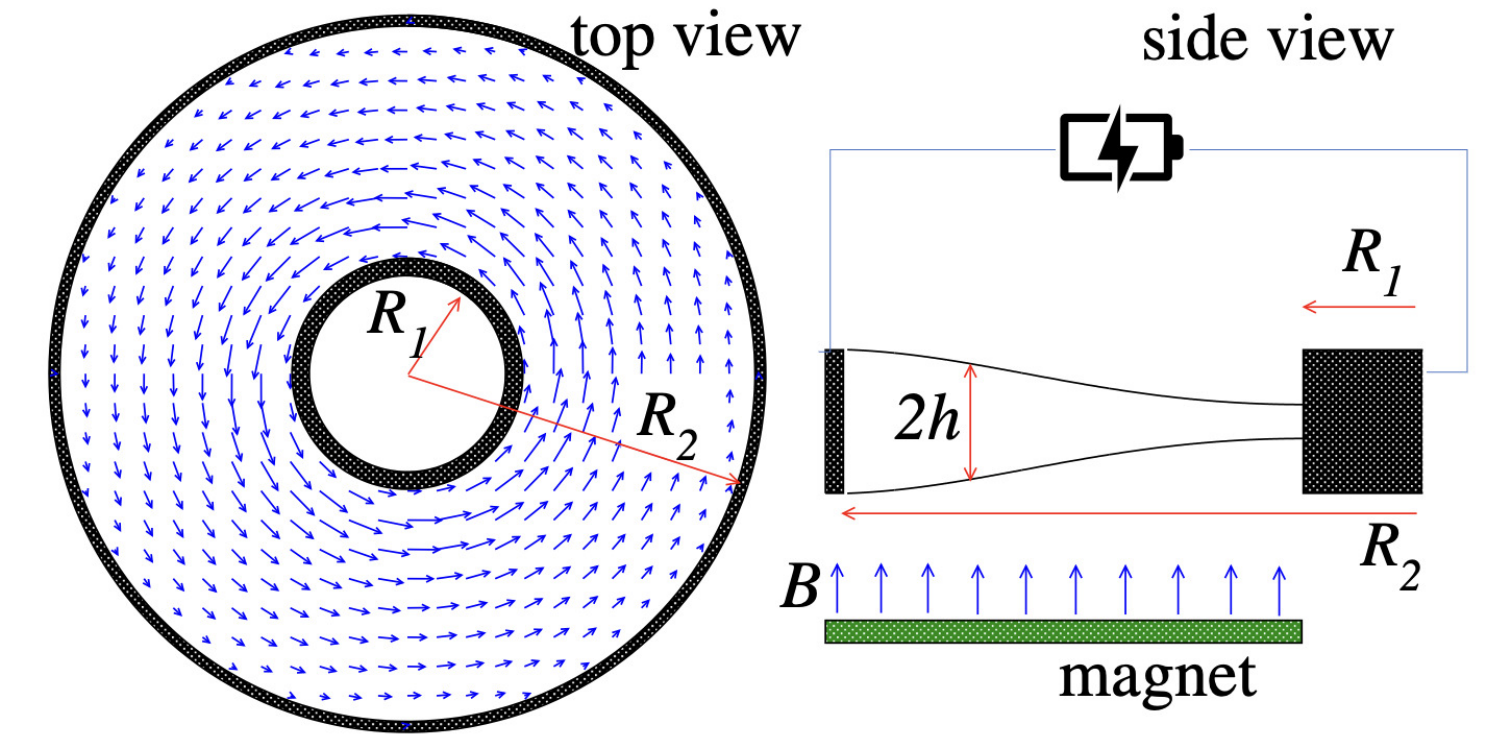}}
  \caption{
  The top (left) and side (right) views of the radial cross-section of
  an annular free film of electrically conducting fluid spanning the
  gap between two coaxial cylindrical electrodes with radii $R_1$ and
  $R_2>R_1$. The film is placed in a vertical uniform magnetic field
  $\bB=(0,0,B)$. Electric current flowing through the film
  between the two electrodes generates Lorentz force that drives the
  flow azimuthally.
  %{\SAS[Perhaps, replace label $B$ with $\tilde\bB$ in the
    %figure to denote a dimensional quantity.]}
    \label{fig1}} 
\end{figure}

%{\SAS [Avoid mixing dimensional and non-dimensional quantities.]}
Inspired by these experiments, we consider an electromagnetically
driven flow in an unsupported free film between two conducting coaxial
cylindrical electrodes with radii $R_1$ and $R_2>R_1$ and placed in a
vertical uniform magnetic field $\bB=(0,0,B)$ 
as schematically shown in figure~\ref{fig1}. 
%{\color{red}{\tt just a note here: I think we are ok to use $B$ for the dimensional magnetic field because it was originally introduced in \ref{eq1} as $B(x,y)$ and then later non-dimensionalised using some reference value $\tB$. All we need to say is that for a homogeneous field $\bB=(0,0,B)$, we choose $\tB=B$ in the definition of the Hartmann number.}}
The potential difference
between the inner and outer electrodes is $V$. 

In what follows we scale the radial coordinate $r$ with $R_2-R_1$, the
film thickness $h$ with its average value $\langle h\rangle$ and
choose the flow velocity scale in such a way that $\Ca\Ree=1$ in
\ref{eq9}, that is
$U=\sqrt{\gamma\langle h\rangle/\left(\rho(R_2-R_1)^2\right)}$. 
With $B$ used as the magnetic field scaling, the Hartmann number becomes  $\Ha^2=B^2(R_2-R_1)^2Kc_{b}^{(0)}/
\mu$, where $c_b^{(0)}$ is the average solute concentration in the bulk.
The
electric potential is scaled with $UB(R_2-R_1)$, the pressure with
$\rho U^2$ and all other dimensionless parameters $\Ma$, $\Ha$, $\Ca$,
$\Pe$ and $\Sc$ are obtained from \ref{eq9} by setting $L=R_2-R_1$.
The dimensionless inner and outer radii are given by $\alpha$ and
$1+\alpha$, respectively, where $\alpha=R_1/(R_2-R_1)$. 
The concentration of a surfactant is scaled
using the average value $c_s^{(0)}$.
Using the same symbols for non-dimensionless fields, we
convert the invariant form of equations~\ref{phys_eq1} to polar
coordinates $(r,\theta)$ and obtain
\begin{eqnarray}
  \partial_tu_r
  &+&u_r\partial_ru_r+\frac{u_\theta}{r}\partial_\theta u_r-\frac{u_\theta^2}{r}
      =g(h)\partial_rh+\partial_r\left(\frac{\partial_rh}{r}
      +\partial_r^2h+\frac{\partial_\theta^2h}{r^2}\right)\nonumber\\
  &+&\Ca\left(\frac{\partial_ru_r}{r}+\partial_r^2u_r
      +\frac{\partial_\theta^2u_r}{r^2}-\frac{u_r}{r^2}
      -\frac{2\partial_\theta u_\theta}{r^2}\right) -\Ca\Ha^2u_r
      -\Ca\Ha^2 \frac{\partial_\theta\phi}{r}+\frac{\Ca}{h}V_r\nonumber\\
  &+&3\Ca\partial_r\left(\frac{u_r}{r}+\partial_r u_r
      +\frac{\partial_\theta u_\theta}{r} \right)
      -\frac{\Ca\Ma}{\Pe}\frac{\partial_rc_s}{h}\,,
\end{eqnarray}
\begin{eqnarray}
  \partial_tu_\theta
  &+& u_r\partial_ru_\theta+\frac{u_\theta}{r}\partial_\theta u_\theta
      +\frac{u_\theta u_r}{r}=\frac{1}{r}g(h)\partial_\theta h
      +\frac{1}{r}\partial_\theta\left(\frac{\partial_r h}{r}+\partial_{r}^2h
      +\frac{\partial_\theta^2h}{r^2}\right)\nonumber\\
  &+&\Ca\left(\frac{\partial_ru_\theta}{r}+\partial_{r}^2u_\theta
      +\frac{\partial_{\theta}^2u_\theta}{r^2}-\frac{u_\theta}{r^2}
      +\frac{2\partial_\theta u_r}{r^2}\right)+\Ca\Ha^2\partial_r \phi
      +\frac{\Ca}{h}V_\theta-\Ca\Ha^2u_\theta\nonumber\\
  &+&\frac{3\Ca}{r}\partial_\theta\left(\frac{u_r}{r}+\partial_r u_r
      +\frac{\partial_\theta u_\theta}{r} \right)
      -\frac{\Ca\Ma}{\Pe}\frac{\partial_\theta c_s}{hr}\,,\label{stat_eq1}
\end{eqnarray}
with 
\begin{eqnarray}
  V_r
  &=&2\left(\partial_rh\,\partial_ru_r
      +\frac{1}{r^2}\partial_\theta h\,\partial_\theta u_r
      -\frac{1}{r^2}u_\theta\,\partial_\theta h\right)
      +\frac{1}{r^2}(\partial_r[ru_\theta]
      -\partial_\theta u_r)\,\partial_\theta h\nonumber\\
  &&+\frac{2}{r}(\partial_r[ru_r]+\partial_\theta u_\theta)\partial_r h\,,\\
  V_\theta
  &=&2\left(\partial_rh\,\partial_ru_\theta
      +\frac{1}{r^2}\partial_\theta h\,\partial_\theta u_\theta
      +\frac{1}{r^2}u_r\,\partial_\theta h\right)
      -\frac{1}{r}(\partial_r[ru_\theta]-\partial_\theta u_r)\,\partial_r h
      \nonumber\\
  &&+\frac{2}{r^2}(\partial_r[ru_r]
     +\partial_\theta u_\theta)\,\partial_\theta h\,.\label{stat_eq2}
\end{eqnarray}
The dynamic equations for the salt and surfactant concentrations and
the kinematic condition are given by
\begin{eqnarray}
  \partial_t[hc_b]
  &+&\frac{1}{r}(\partial_r[rhc_bu_r]+\partial_\theta[hc_bu_\theta])
      =\frac{\Sc^{-1}\Ree^{-1}}{r^2}[r\partial_r[rh\partial_rc_b]
      +\partial_\theta[h\partial_\theta c_b]]\,,\\
  \partial_t c_s
  &+&\frac{1}{r}(\partial_r[rc_su_r]+\partial_\theta[c_su_\theta])
      =\frac{\Pe^{-1}}{r^2}[r\partial_r[r\partial_rc_b]
      +\partial_\theta^2c_s]\,,\\
  \partial_th
  &+&\frac{1}{r}(\partial_r[rhu_r]+\partial_\theta[hu_\theta])=0\,.
      \label{stat_eq3}
\end{eqnarray}
The system is completed by the continuity equation for the electric
current
\begin{equation}
  \partial_r[c_brh(u_\theta-\partial_r\phi)]
  -\partial_\theta\left[\frac{c_bh}{r}(ru_r+\partial_\theta\phi)\right]
  =0\,.\label{stat_eq4}
\end{equation}
%The scaled disjoining pressure is given by 
%$\psi=\frac{A_s}{ (2h)^3}- E_s e^{-2\kappa_s h}$ with the
%dimensionless Hamaker constant $A_s=A/(6\pi \rho U^2\langle
%h\rangle^3)$, eccess energy factor $E_s=\kappa E/(\rho U^2)$ and
%$\kappa_s=\kappa \langle h\rangle$.

System of equations~\ref{stat_eq1}--\ref{stat_eq4} admits a steady
solution that corresponds to the azimuthal flow field $u_r=0$,
$u_\theta=f(r)$ induced by axisymmetric electric potential $\phi(r)$
in a film with axisymmetric profile $h(r)$ containing uniformly
dissolved salt with constant bulk concentration $c_b=1$ and
covered by a uniformly distributed surfactant with surface
concentration $c_s=1$. The functions  $f(r)$, $h(r)$ and
$\phi(r)$ are found from
\begin{eqnarray}
  &\left(\frac{(rh')'}{r}\right)'+g(h)h'+\frac{f^2}{r}=0\,,&\label{stat_eq5a}\\
  &f''+\frac{f'}{r}-\frac{f}{r^2}-\Ha^2f+\Ha^2\phi'
  +\frac{h'}{h}\left(f'-\frac{f}{r}\right)=0\,,&\label{stat_eq5b}\\
  &(r h(\phi'-f))'=0\,,&\label{stat_eq5c}
\end{eqnarray}
where primes denote the radial derivative $\frac{\dd}{\dr}$.

Equation~\ref{stat_eq5c} is integrated once to yield
\begin{equation}
  \phi'=f-\frac{e}{rh}\,,\label{stat_eq6}
\end{equation}
where constant $e$ is linked to the potential difference
$\phi(1+\alpha )-\phi(\alpha)=\Delta\phi$ between the electrodes via 
\begin{equation}
  \Delta\phi=\int_\alpha^{1+\alpha}f(r)\dr-e
  \int_\alpha^{1+\alpha}\frac{\dr}{rh}\,.\label{stat_eq6a}
\end{equation}
 The term $e/(rh)$ in equation~\ref{stat_eq6}
represents the radial current density
\begin{equation}
  j_r=f-\phi'=\frac{e}{rh}\,.\label{stat_curr}
\end{equation}
The total steady current through the vertical cylindrical
section of the film at any radial location $\alpha\leq r\leq1+\alpha$
is independent of the radius of a cross-section and is given by $4\pi
rhj_r=4\pi e$.

%Note that the expression Eq.\,(\ref{stat_eq6}) is identical to the
%radial gradient of the electric potential $\phi'$ in the thin layer
%approximation found for a flat electrically conducting fluid layer
%support by a solid substrate with no-slip boundary
%\cite{Perez16,Suslov17}. 

%The first equation in equations\,(\ref{stat_eq5})  represents the
%balance between the centrifugal force and the radial component of the
%gradient of the Laplace and disjoining pressure.  

Eliminating $\phi$ from equation~\ref{stat_eq5b} we obtain two
coupled equations for $f$ and $h$
\begin{eqnarray}
  &\left(\frac{(rh')'}{r}\right)'+g(h)h'+\frac{f^2}{r}=0\,,&\label{stat_eq7a}\\
  &f''+\frac{f'}{r}-\frac{f}{r^2}-\frac{\Ha^2e}{rh}
       +\frac{h'}{h}\left(f'-\frac{f}{r}\right)=0&\,.\label{stat_eq7b} 
\end{eqnarray}
Fluid velocity vanishes at the surface of the electrodes leading to
$f(\alpha)=f(1+\alpha )=0$. The boundary condition for the film
deformation must be compatible with the long-wave approximation used
here, which only takes into account relatively small film slopes
$h'\ll1$. In what follows we assume $h'(\alpha)=h'(1+\alpha )=0$.

We are looking for a solution of the boundary value
problems~\ref{stat_eq7a} and \ref{stat_eq7b} that corresponds to
a given average film half-thickness
\begin{equation}
  \frac{2}{1+2\alpha}\int_{\alpha}^{1+\alpha }h(r)r\dr=1
  \label{h2}
\end{equation}
and satisfies additional integral condition~\ref{stat_eq6a} for any
given value of the applied voltage $\Delta\phi$. 

For reference, we derive the approximate analytic solution that
corresponds to the flow in a flat undeformed film. By setting $h=1$
and neglecting the centrifugal acceleration $f^2/r$ we find 
from equations~\ref{stat_eq7b} and \ref{stat_eq6a}
\begin{equation}
  f=\alpha\Ha^2e
  \left[C\left(\frac{\alpha}{r}-\frac{r}{\alpha}\right)
    +\frac{r}{2\alpha}\ln\frac{r}{\alpha}\right]\,,\label{stat_eq8}
\end{equation}
where 
$C=\frac{(1+\alpha)^2\ln{(1+\alpha^{-1})}}{2(1+2\alpha)}$,
$e=\frac{\Delta \phi}{\alpha^2\Ha^2D-\ln(1+\alpha^{-1})}$ and
$D=C\ln\left(1+\alpha^{-1}\right)-\frac{1+2\alpha}{8\alpha^2}$.

To find non-trivial solutions of the boundary value problem
\ref{stat_eq7a}, \ref{stat_eq7b}  with the integral condition
\ref{h2} we use a numerical continuation package AUTO
\citep{Krauskopf14, AUTO94}. The trivial solution $f=0$ and $h=1$ that
exists for $\Delta\phi=0$ is used as a starting point for numerical
continuation with $\Delta\phi$ being gradually increased.

In the absence of the disjoining pressure, that is for $g(h)=0$, the
boundary value problem \ref{stat_eq7a}, \ref{stat_eq7b} has an
important scaling property. Namely, it contains three dimensionless
parameters, $\Ha$, $\alpha$ and $\Delta\phi$,
but only one of them, $\Delta\phi$, depends
explicitly on the dimensional layer thickness $2\langle h\rangle$ via
$V=\Delta\phi B\sqrt{\gamma\langle h\rangle/\rho}$.
%$\Ha=B(R_2-R_1)\sqrt{Kc_b^{(0)}/\mu}$,  $\alpha=R_1/(R_2-R_1)$} and  
%the voltage $\Delta\phi$, where $V$ is the applied dimensional voltage. 
This implies that for any fixed $\Ha$ and $\alpha$ there exists a
universal branch of solutions parameterised by $\Delta\phi$. A
solution that corresponds to an arbitrary value of the average film
half-thickness $\langle h\rangle$ and an arbitrary applied 
voltage $V$ is found on the universal branch for
$\Delta\phi=(V/B)\sqrt{\rho/(\gamma\langle h\rangle)}$.

\begin{figure} 
  \centerline{\includegraphics[width=\columnwidth]{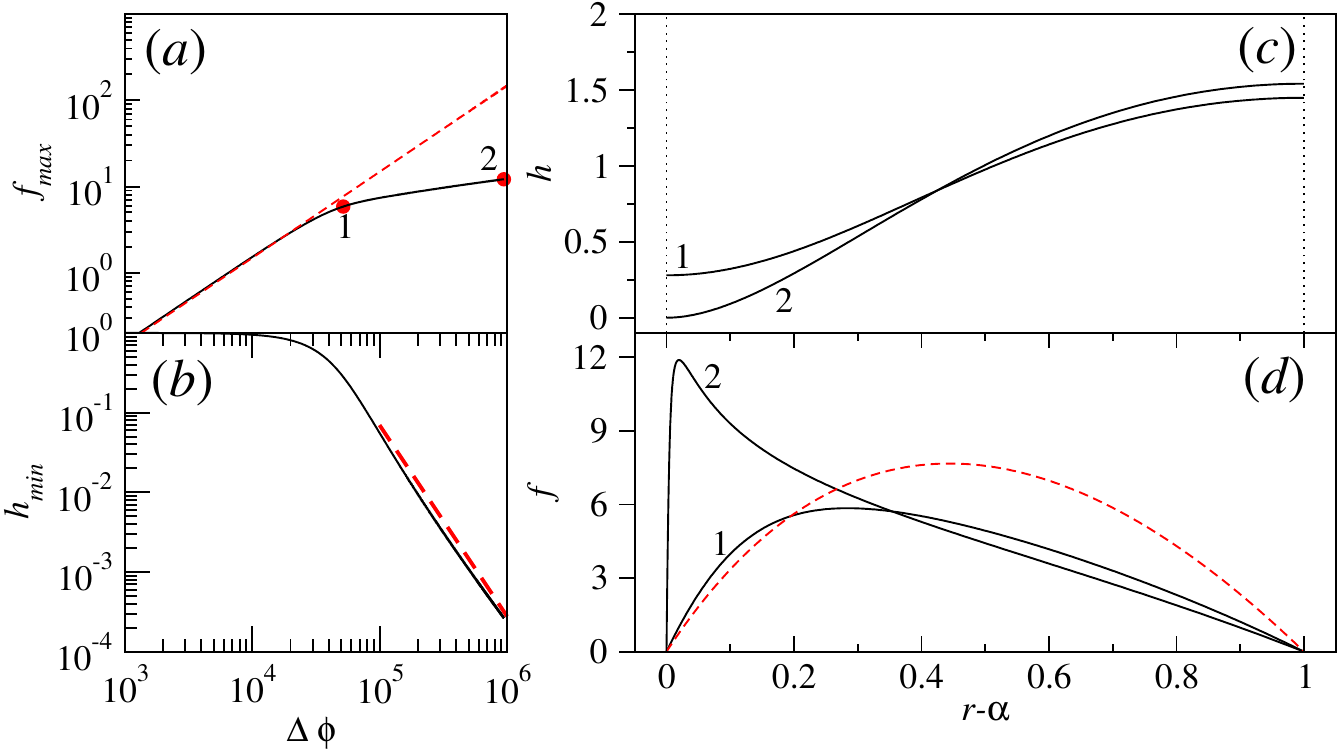}}
  \caption{($a$) Maximum fluid velocity in a free annular film as a
    function of the applied voltage $\Delta\phi$ for
    $\Ha^2=1.3\times10^{-3}$ and $\alpha=1$. The dashed and solid
    lines correspond to the flat film approximation \ref{stat_eq8}
    and the numerical solution, respectively. ($b$) Minimum film
    thickness $h_{\min}$ at the inner cylinder as a function of the
    applied voltage. The dashed line depicts the power law function
    $\sim(\Delta \phi)^{-2.4}$. ($c$) Film thickness $h(r)$ for the
    values of $\Delta\phi$ at points $1$ and $2$ labelled in panel
    ($a$). ($d$) Velocity $f(r)$ for solutions at points $1$ and $2$
    in panel ($a$) (solid lines) and corresponding to the flow in the
    flat-film approximation \ref{stat_eq8} (dashed
    lines).\label{fig2}}  
\end{figure}

Taking into account the above scaling property we consider a free
film with an arbitrary average half-thickness $\langle h\rangle$
spanning the gap between cylindrical electrodes with radii $R_1=1$\,cm
and $R_2=2$\,cm ($\alpha=1$). As an example we chose fluid properties
and magnetic field strength similar to those used in \cite{Perez16, Suslov17}:
$Kc_b^{(0)}=\sigma=5$\,Ohm\,m$^{-1}$, $\mu=0.001$\,kg\,m\,s$^{-1}$,
$\rho=1000$\,kg\,m$^{-3}$, $B=0.05$\,T. Additionally, we use
$\gamma=0.03$\,N\,m$^{-1}$ for the surface tension coefficient. These
correspond to $\Ha^2\approx1.3\times10^{-3}$. As the solution measure
we take the maximum flow velocity $f_{\max}$ and the minimum film
thickness $h_{\min}$ attained at the inner cylinder ($r=\alpha$). The
numerically obtained maximum velocity $f_{\max}$ is shown in
figure~\ref{fig2}($a$) by the solid line while the value found from
the flat-film approximation \ref{stat_eq8} is depicted by the dashed
line. The minimum film thickness is shown in the log-log scale in
figure~\ref{fig2}($b$) indicating that $h_{\min}$ asymptotically
approaches zero as a power law function
$\sim(\Delta\phi)^{-2.4}$. This implies that the solution exists for
any fixed value of $\Delta\phi$ no matter how large it is. However,
the film thickness becomes vanishingly small at the inner electrode
indicating that the film is likely to rapture there. The film profile
and the flow velocity at points $1$ and $2$ are shown in
figures~\ref{fig2}($c$) and \ref{fig2}($d$), respectively. The dashed 
line in figure~\ref{fig2}($d$) corresponds to the approximate
solution~\ref{stat_eq8}. 

\section{Linear stability of the azimuthal flow}\label{stab}
In this section, we study the linear stability of the base azimuthal
flow $(u_r,u_\theta)=(0,f(r))$ in a free film with the half-thickness
$h_0(r)$ in the absence of the disjoining pressure ($\Pi(h)=0$). It is
anticipated that the least stable perturbations are azimuthally
invariant due to the stabilising effect of the surface
tension. Indeed, any perturbation that varies azimuthally must 
be periodic in $\theta$ and, consequently, be proportional to $e^{\ii
  n\theta}$, $n=0,1,2\ldots$. Therefore, the magnitude of the
stabilising surface tension terms in equation~\ref{stat_eq1}
increases as $n^3$ and is the smallest for $n=0$. 

Equations~\ref{stat_eq1}--\ref{stat_eq4} are linearised about the
base flow by writing
\begin{eqnarray}
  &u_r=e^{\lambda t}\tu_r(r)\,,\quad
    u_\theta=f(r)+e^{\lambda t}\tu_\theta(r)\,,\quad
    h=h_0+e^{\lambda t}\thh\,,&\nonumber\\
  &c_b=1+e^{\lambda t}\tc_b\,,\quad
    c_s=1+e^{\lambda t}\tc_s\,,\quad
    \phi'=f-\frac{e}{rh_0}+e^{\lambda t}\tphi'(r)\,,&\label{sub}
\end{eqnarray}
where $\lambda$ is the perturbation growth rate, $h_0=h_0(r)$ is the
steady film profile and the tilded variables represent small-amplitude
perturbations, substituting these in the equations and neglecting the
products of perturbations. After dropping the tildas we obtain
\begin{eqnarray}
  \lambda u_r
  &=&-\frac{2fu_\theta}{r}
      +\left(\frac{(rh')'}{r}\right)'+4\Ca\left(\frac{(ru_r)'}{r}\right)'-\Ca\Ha^2u_r\nonumber\\
      &&+\frac{2\Ca h_0'}{h_0}\left(\frac{u_r}{r}+2u_r'\right)
  -\frac{\Ca\Ma}{\Pe h_0} c'_s\,,\label{ls_eq1a}\\
  \lambda u_\theta 
  &=&-u_rf'-\frac{fu_r}{r}+\Ca\left(\frac{(ru_\theta)'}{r}\right)'
      -\Ca\Ha^2\left(u_\theta-\phi'\right)\nonumber\\
      &&+\frac{\Ca h_0'}{h_0}\left(u_\theta'-\frac{u_\theta}{r}\right)
  +\Ca\left(\frac{h}{h_0}\right)'\left(f'-\frac{f}{r}\right)\,,\label{ls_eq1b}\\
  \lambda h&=&-\frac{1}{r}\left(rh_0u_r\right)'\,,\label{ls_eq1c}\\
  \lambda c_bh_0+\lambda h&=&-\frac{1}{r}(rh_0u_r)'
       +\Sc^{-1}\Ree^{-1}\frac{(rh_0(c_b)')'}{r}\,,\label{ls_eq1d}\\
  \lambda c_s&=&-\frac{1}{r}\left(ru_r\right)'
       +\Pe^{-1}\frac{(r(c_s)')'}{r}\,,\label{ls_eq1e}\\
  0&=&\left(rh_0(u_\theta-\phi')\right)'
       +\left(\frac{e\left( h+c_b h_0\right)}{ h_0}\right)'\,.\label{ls_eq1f}
\end{eqnarray}
The flow velocity vanishes at $r=\alpha$ and $r=1+\alpha$ so that
$u_r=u_\theta=0$ there. This leads to the automatic conservation of
the total volume of the fluid 
\begin{equation}
  \int_\alpha^{1+\alpha}2\pi hr\dr=
  -\frac{1}{\lambda}\int_\alpha^{1+\alpha}
  \frac{2\pi}{r}\left(rh_0u_r\right)'r\dr=0\,.\label{ls_eq2}
\end{equation}
Additionally, we require $h'(\alpha)=h'(1+\alpha)=0$, which with the
help of~\ref{ls_eq1c} and conditions $h_0'(\alpha)=h_0'(1+\alpha)=0$
and $u_r(\alpha)=u_\theta(\alpha)=u_r(1+\alpha)=u_\theta(1+\alpha)=0$
translates into two boundary conditions for the radial velocity $u_r$ 
\begin{equation}
  u_r'(\alpha)+u_r''(\alpha)
  =\frac{u_r'(1+\alpha)}{1+\alpha}+u_r''(1+\alpha)=0\,.\label{ls_eq2a}
\end{equation}

Integrating the equation for the perturbation of the electric
potential \ref{ls_eq1f} once we obtain
\begin{equation}
  u_\theta-\phi'=\frac{c}{rh_0}-\frac{eh}{rh_0^2}
  -\frac{ec_b}{rh_0}\,,\label{ls_eq3}
\end{equation}
where $c$ is some constant. Integrating \ref{ls_eq3} and imposing
the condition $\phi(\alpha)=\phi(1+\alpha)=0$ we find 
\begin{equation}
  \int_\alpha^{1+\alpha}u_\theta\dr=c\int_\alpha^{1+\alpha}\frac{\dr}{rh_0}
  -e\int_\alpha^{1+\alpha}\frac{h}{rh_0^2}\dr
  -e\int_\alpha^{1+\alpha}\frac{c_b}{rh_0}\dr\,.\label{ls_eq4}
\end{equation}
Multiplying equations~\ref{ls_eq1a} and \ref{ls_eq1b} by
$\lambda$, using equations~\ref{ls_eq1c} and \ref{ls_eq3} to
eliminate $h$ and $\phi'$ and, subsequently, differentiating
\ref{ls_eq1d} and \ref{ls_eq1e} with respect to the radius $r$ we
arrive at a nonlinear eigenvalue problem 
\begin{eqnarray}
  \lambda^2 u_r-\frac{2\lambda fu_\theta}{r}
  &=&-\left(\frac{(r(r^{-1}\left(rh_0u_r\right)')')'}{r}\right)'
      +4\lambda\Ca\left(\frac{(ru_r)'}{r}\right)'
      -\lambda \Ca\Ha^2u_r\nonumber\\
  &&+\frac{2\Ca\lambda h_0'}{h_0}\left(\frac{u_r}{r}+2u_r'\right)
     -\frac{\Ca\Ma}{\Pe}\frac{\lambda c_s'}{h_0}\,,\label{ls_eq5a}\\
  \lambda^2 u_\theta+\lambda u_rf'
  &=&-\frac{\lambda fu_r}{r}
      +\lambda\Ca\left(\frac{(ru_\theta)'}{r}\right)'
      -\Ca\Ha^2\left(\frac{e (rh_0 u_r)'}{r^2h_0^2}
      -\frac{\lambda e c_b}{rh_0}
      +\frac{\lambda c}{rh_0}\right)\nonumber\\
  &&+\frac{\lambda\Ca h_0'}{h_0}
     \left(u_\theta'-\frac{u_\theta}{r}\right)
     -\Ca\left(\frac{(rh_0u_r)'}{rh_0}\right)'\left(f'-\frac{f}{r}\right)\,,
     \nonumber\\
  \lambda c_b'
  &=&\Sc^{-1}\Ree^{-1}\left(\frac{(rh_0c_b')'}{rh_0}\right)'\,,\label{ls_eig}\\
  \lambda c_s'
  &=&
      \Pe^{-1}\left(\frac{(rc_s')'}{r}\right)'
      -\left(\frac{\left(ru_r\right)'}{r}\right)'\label{ls_eq5}
\end{eqnarray}
that must be solved in conjunction with the integral condition
\ref{ls_eq4}. The set of boundary conditions \ref{ls_eq2a} and
$u_r(\alpha)=u_\theta(\alpha)=u_r(1+\alpha)=u_\theta(1+\alpha)=0$ must
be extended with 
\begin{eqnarray}
  c_s'(\alpha)=c_b'(\alpha)=c_s'(1+\alpha)=c_b'(1+\alpha)=0\,,\label{ls_bc}
\end{eqnarray}
which accounts for the chemically passive impenetrable boundaries of
the two electrodes. Note that unlike \ref{stat_eq7a} and
\ref{stat_eq7b} the eigenvalue problem \ref{ls_eq5} contains the
average film thickness $\langle h\rangle$ as a part of the capillary
number $\Ca$.

\subsection{Linear stability of a flat film with no applied
  voltage}\label{stab1}
In this section, we consider the stability of a flat film in the
absence of an electric current. With $h_0=1$, $f=0$ and $e=0$ the
eigenvalue problem \ref{ls_eq5a}--\ref{ls_eq5} and the integral
condition \ref{ls_eq4} reduce to three decoupled eigenvalue
problems: one for the azimuthal velocity perturbations $u_\theta$, one
for the radial velocity $u_r$ and surfactant $c_s$ perturbations and
one for the perturbation of the solute concentration $c_b$: 
\begin{eqnarray}
  \lambda u_\theta 
  &=&\Ca\oL[u_\theta]
      -\frac{\Ca\Ha^2}{r\ln\left(1+\alpha^{-1}\right)}
      \int_\alpha^{1+\alpha}u_\theta\dr\,,\label{flat2}\\
  \lambda^2 u_r
  &=&- \oLL[u_r]+4\lambda\Ca\oL[u_r]
      -\lambda\Ca\Ha^2u_r-\Ca\Ma\Pe^{-1}\lambda c_s'\,,\label{flat1}\\
  \lambda c_s'&=&\Pe^{-1}\oL[c_s'] -\oL[u_r]\,,\label{flat4}\\
  \lambda c_b'&=&\Sc^{-1}\Ree^{-1}\oL[c_b']\,,\label{flat3}
\end{eqnarray}
where $\oL[u]\equiv\left(r^{-1}(ru)'\right)'$.

The eigenvalue problem~\ref{flat2}--\ref{flat3} can be solved
analytically in terms of the auxiliary eigenvalue problem for the
operator $\cL$ 
\begin{equation}
  \oL[u]=-\Lambda u\,,\label{op1}
\end{equation}
which coincides with Bessel's differential equation. The solution of
\ref{op1} that satisfies the Dirichlet boundary conditions 
$u(\alpha)=u(1+\alpha)=0$ exists only for real positive $\Lambda$
and is given by
\begin{equation}
  u(r)=C_1 J_1\left(r\sqrt{\Lambda}\right)
  +C_2Y_1\left(r\sqrt{\Lambda}\right)\,,\label{op2}
\end{equation}
where $C_1$ and $C_2$ are arbitrary constants and $J_1$ and $Y_1$ are
the Bessel functions of order one of the first and second kind,
respectively. Applying the boundary conditions we obtain the
solvability condition for constants $C_1$ and $C_2$ that determines
the entire spectrum of discrete eigenvalues $\Lambda$ 
\begin{equation}
  J_1\left(\alpha\sqrt{\Lambda}\right)
  Y_1\left((1+\alpha)\sqrt{\Lambda }\right)
  -J_1\left((1+\alpha)\sqrt{\Lambda}\right)
  Y_1\left(\alpha\sqrt{\Lambda}\right)=0\,.\label{op3}
\end{equation}

It follows from equations \ref{op1}--\ref{op3} that the spectrum
of eigenvalues $\lambda_{c_b}$ of the bulk solute concentration
perturbations \ref{flat3} is real and negative 
\begin{equation}
  \lambda_{c_b}=-(\Sc\Ree)^{-1}\Lambda\label{cb_eq1}
\end{equation}
and the corresponding eigenfunction is given by
\begin{equation}
  c_b'(r)=C\left[ J_1\left(r\sqrt{\Lambda}\right)
    Y_1\left(\alpha\sqrt{\Lambda}\right)
    -Y_1\left(r\sqrt{\Lambda}\right)
    J_1\left(\alpha\sqrt{\Lambda}\right)\right]\,,\label{cb_eq2}
\end{equation}
where $C$ is an arbitrary constant.

The eigenvalue problem \ref{flat1} and \ref{flat4} for $u_r$ and
$c_s'$ can be solved in a similar way using the eigenfunctions of the
operator $\cL$. The solution that satisfies
$u_r(\alpha)=u_r(1+\alpha)=0$ and $c_s'(\alpha)=c_s'(1+\alpha)=0$ is
given by
\begin{eqnarray}
  u_r(r)
  &=&C_1\left[ J_1\left(r\sqrt{\Lambda}\right)
      Y_1\left(\alpha\sqrt{\Lambda}\right)
      -Y_1\left(r\sqrt{\Lambda}\right)
      J_1\left(\alpha\sqrt{\Lambda}\right)\right]\,,\nonumber\\
  c_s'(r)
  &=&C_2u_r\,,\label{ls_eq7}
\end{eqnarray}
where $C_1$ and $C_2$ are some constants.

Substituting \ref{ls_eq7} into \ref{flat1} and \ref{flat4} we obtain 
\begin{eqnarray}
  \lambda^2
  &=&-\Lambda^2-\Ca(4\Lambda+\Ha^2)\lambda
      -C_2\Ca\Ma\Pe^{-1}\lambda\,,\label{ls_eq9}\\
  \lambda C_2
  &=&\Lambda-\Pe^{-1}\Lambda C_2\label{ls_eq90}
\end{eqnarray}
and then eliminating $C_2$ from \ref{ls_eq9} we arrive at a cubic
equation for $\lambda$
\begin{equation}
  (\Pe\lambda+\Lambda)\left(\lambda^2
    +\Ca\left(4\Lambda+\Ha^2\right)\lambda+\Lambda^2\right)
  +b\Lambda\lambda\Pe=0\,,\label{eig_ur}
\end{equation}
where $b=\Ca\Ma\Pe^{-1}>0$ characterises the strength of the Marangoni
flow.

For any admissible value of $\Lambda$ from \ref{op3} one needs to
solve \ref{eig_ur} to find the eigenvalue $\lambda$. Then the
corresponding eigenfunctions~\ref{ls_eq7} only contain one arbitrary
scaling factor $C_1$ with $C_2=\Lambda/(\lambda+\Pe^{-1}\Lambda)$. 
Here we consider two physically distinct situations:  the 
diffusion-dominated regime, when the surface diffusivity is large,
i.e.~$d_s\to\infty$, and the advection-dominated regime, when $d_s\to
0$. Since the Hartmann number $\Ha$, the capillary number $\Ca$ and
parameter $b=\Ca\Ma\Pe^{-1}$ in equation~\ref{eig_ur} do not depend
on $d_s$, the diffusion-dominated regime corresponds to $\Pe\to0$
with $\Ha$, $\Ca$ and $b$ remaining finite. In this case
equation~\ref{eig_ur} has three distinct solutions
\begin{eqnarray}
  \lambda_{1,2}
  &=&
      -\frac{\Ca\left(4\Lambda+\Ha^2\right)+b\Pe}{2}
      \pm\frac{\Omega}{2}\left(1+b\Pe\frac{\Ca(4\Lambda+\Ha^2)}
      {2\Omega^2}\right)+O(\Pe^2)\,,\label{ls_eq10a}\\
  \lambda_3
  &=&-\Pe^{-1}\Lambda+b\Pe+O(\Pe^2)\,,\label{ls_eq10b}
\end{eqnarray}
where $\Omega=\sqrt{\Ca^2(4\Lambda+\Ha^2)^2-4\Lambda^2}$.

Note that $\lambda_3$ is real and negative and its magnitude is always
much larger than $|\lambda_{1,2}|$. Since $\Lambda$ is real and
positive, we conclude that the real parts of $\lambda_{1,2}$ are
always negative. Moreover, $\lambda_{1,2}$ become complex if
$\Ca\left(4\Lambda+\Ha^2\right)<2\Lambda$ implying that the radial
perturbation mode undergoes the transition from a monotonic to
oscillatory decay. In physical variables, the condition for the
oscillatory decay of the radial mode is 
\begin{equation}
  \sqrt{\rho\gamma\langle h\rangle}>
  2\mu+\frac{B^2 (R_2-R_1)^2\sigma}{2\Lambda}\,,\label{ls_osc}
\end{equation}
where the electric conductivity is $\sigma=Kc_b^{(0)}$ and $\Lambda$
depends on the ratio of the radii $R_2/R_1$ via parameter $\alpha$.
It is instructive to compare condition~\ref{ls_osc} with the linear
stability of an unbounded flat free layer in the absence of the
magnetic field. Neglecting the disjoining pressure, the growth rate
$\omega(k)$ of the least stable mode with the wave vector $k$ in an
unbounded horizontal free film can be obtained from equation~(30) in
\cite{Erneux93}: 
\begin{equation}
  \frac{\mu}{\rho}\omega(k)=-2k^2
  +k^2\sqrt{4-\frac{\rho\langle h\rangle\gamma}{\mu^2}}\,.
  \label{ls_osc2} 
\end{equation}
It follows from \ref{ls_osc2} that the critical thickness
$2\langle h\rangle_c$ of the layer above which the relaxation dynamics
is oscillatory is given by 
\begin{equation}
  2\langle h\rangle_c=\frac{8\mu^2}{\rho\gamma}\,.\label{ls_osc3}
\end{equation}
This coincides with our result~\ref{ls_osc} for $B=0$.

It is seen from equations~\ref{ls_eq10a} that in the absence of the
Marangoni flow, that is when $b=0$, the eigenvalue of the radial
velocity perturbation with the largest real part is given by 
\begin{equation}
  \lambda_1=-\frac{\Ca\left(4\Lambda+\Ha^2\right)}{2}
  +\frac{\Omega}{2}\,.\label{zero_cs}
\end{equation}
The perturbation of the surfactant is decoupled from that of the
radial flow and has a negative real eigenvalue $\lambda_3=-\Lambda
\Pe^{-1}$. In this regime, any perturbation of the surfactant
distribution relaxes on the time scale $|\lambda_3|^{-1}$, which is
much shorter than the characteristic decay time
$-\Re({\lambda_{1,2}}^{-1})$ of fluid motion. In the presence of the
Marangoni flow when $b\ne0$, the perturbations of the radial velocity
and surfactant concentrations are coupled and their dynamics is
characterised by the leading eigenvalues $\lambda_{1,2}$. The
Marangoni flow leads to a further stabilisation of the leading
perturbation mode as follows from equation~\ref{ls_eq10a}. Indeed,
by analysing the real parts of the leading eigenvalues we conclude that
\[\Re(\lambda_{1,2})|_{b>0}<\Re(\lambda_{1,2})|_{b=0}<0\]
regardless of the sign of $\Omega^2$. As a consequence, the
characteristic decay time of the flow perturbation decreases in the
presence of Marangoni effect.

As follows from equation~\ref{flat4}, in the advection-dominated
regime $\Pe\to\infty$ the dynamics of the surfactant perturbation is
governed by the radial flow $u_r$. Indeed in this limit $\lambda
c_s'=-\oL[u_r]$, which shows that the surfactant plays the role of an
active scalar field advected by the flow while its gradient influences
the stability of the flow. By letting $\Pe\to\infty$ in
\ref{ls_eq90} and then substituting $C_2=\Lambda/\lambda$ in
equation~\ref{ls_eq9} we obtain
\begin{eqnarray}
  \lambda_{\pm}
  &\approx&\frac{1}{2}\left[-\Ca\left(4\Lambda+\Ha^2\right)
            \pm\sqrt{\Ca^2\left(4\Lambda+\Ha^2\right)^2
            -4\Lambda(\Lambda+b)}\right]\,,\label{eig_ur1}
\end{eqnarray}
From \ref{eig_ur1} we see that the Marangoni flow has no effect on
the stability of the base flow if the expression under the radical is
negative. However, if 
\[\Ca^2\left(4\Lambda+\Ha^2\right)^2-4\Lambda(\Lambda+b)>0\,,\]
the leading eigenvalue $\lambda_+$ becomes real and negative, and the
presence of surfactant has a stabilising effect since
$|\lambda_+|_{b>0}>|\lambda_+|_{b=0}$. These analytical results extend
an earlier study on the linear stability of planar soap films
\citep{Miksis98} by including the Lorentz force effects. The
observation of the stabilising role of the Marangoni flow in the limit
of diffusion-dominated and advection-dominated regimes is in agreement
with \cite{Miksis98}, where it was also found that for long-wavelength
perturbations the presence of the Marangoni flow decreases the growth
rate of the dominant perturbation mode. 

\begin{figure} 
  \centerline{\includegraphics[width=\columnwidth]{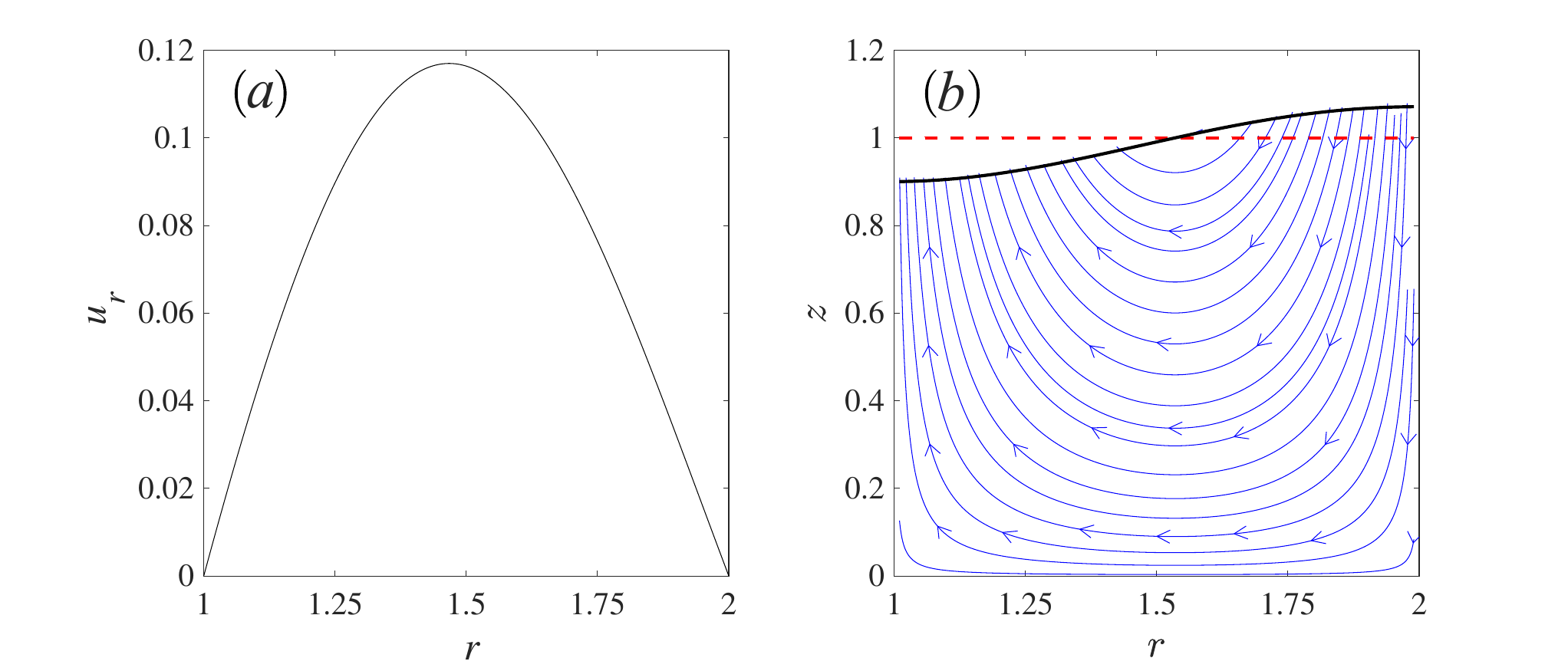}}
  \caption{($a$) Leading eigenfunction
    $u_r=Y_1\left(\alpha\sqrt{\Lambda}\right)
    J_1\left(r\sqrt{\Lambda}\right)
    -J_1\left(\alpha\sqrt{\Lambda}\right)
    Y_1\left(r\sqrt{\Lambda}\right)$
    of the radial perturbation mode for the same parameters as in
    figure~\ref{fig2}. ($b$) Streamlines of the radially perturbed
    flow field in a vertical crosssection of the film.}\label{fig3}
\end{figure}

To illustrate the structure of the oscillatory radial mode we consider
the diffusion-dominated regime in a film with the average thickness
$2\langle h\rangle=20\,\mu$m and take all other parameters as in
figure~\ref{fig2}. For $\alpha=1$, the leading eigenvalue of the operator
$\cL$ is $\Lambda\approx10.218$, which corresponds to the complex
leading eigenvalue of the radial mode
$\lambda_r\approx-1.179\pm10.149\ii$. The corresponding eigenfunction
$u_r$ is shown in figure~\ref{fig3}($a$). 
Figure~\ref{fig3}($b$) depicts the instantaneous streamlines of the
corresponding flow field in a vertical cross-section of the film
$(r,z)$. The streamlines are obtained by recalling that the vertical
flow velocity $w$ is given by $w(r,z)=-r^{-1}(ru_r)'z$ so that the
kinematic equation \ref{ls_eq1c} can be written in the form
$\partial_t h =-w(r,1)=- r^{-1}(ru_r)'$. The film interface oscillates
about $h_0=1$ with a decreasing amplitude while the fluid flows from
the inner to the outer cylinder and back.

The eigenvalue problem (\ref{flat2}) for the azimuthal
velocity perturbation $u_\theta$ can only be found analytically for a weak
magnetic field at $\Ha\to0$. When $\Ha=0$, the spectrum of the
azimuthal velocity perturbations is real and is given by  
\begin{equation}
  \lambda_{u_\theta}=-\Lambda\Ca\,.\label{ls_eq11a}
\end{equation}
Comparing \ref{ls_eq11a} with \ref{ls_eq10a} we observe that in
the absence of the Marangoni flow ($b=0$) and when $\Ha=0$ the
absolute value of the eigenvalue $|\lambda_{u_\theta}|$ of the
monotonically stable azimuthal velocity perturbation is exactly half
of the real part of the eigenvalue $\lambda_{u_r}$ of the radial
velocity mode.

\subsection{Linear stability of a deformed film in the presence of
  electric current}\label{stab2}
In this section, we study the linear stability of the azimuthal flow
in a deformed film in the diffusion-dominated regime, when
perturbations of the surfactant and solute fields relax
instantaneously. The azimuthal and radial velocity perturbation modes
discussed in the previous section become coupled in the presence of an
electric current. This implies that for any infinitesimal applied
voltage $\Delta\phi$ the spectrum of the generalised eigenvalue
problem \ref{ls_eq5} is discrete and contains the eigenvalues that
originate from each of the possible radial and azimuthal velocity
modes existing in the absence of current. 

To visualise how the stability of the azimuthal flow changes with the
applied voltage we choose $\langle h\rangle=10\,\mu$m and keep all
other parameters as in figure~\ref{fig2}. The numerical continuation
method is then employed to track each mode and the corresponding
eigenvalue using $\Delta\phi$ as the continuation parameter. To this
end, the generalised eigenvalue problem \ref{ls_eq5} is solved
simultaneously with equations~\ref{stat_eq7a} and\ref{stat_eq7b}.
To quantify the relative coupling strength between the perturbations of the radial and the
azimuthal velocity we introduce the parameter
\begin{equation}
  \chi=\arctan{\left(\frac{\int_\alpha^{1+\alpha}|u_r|^2\dr}
      {\int_\alpha^{1+\alpha}|u_\theta|^2\dr}\right)}\,.\label{ls_eq12}
\end{equation}
Thus, $\chi=\pi/2$ corresponds to the pure radial and $\chi=0$ to pure
azimuthal modes, respectively.
 
\begin{figure} 
  \centerline{\includegraphics[width=\columnwidth]{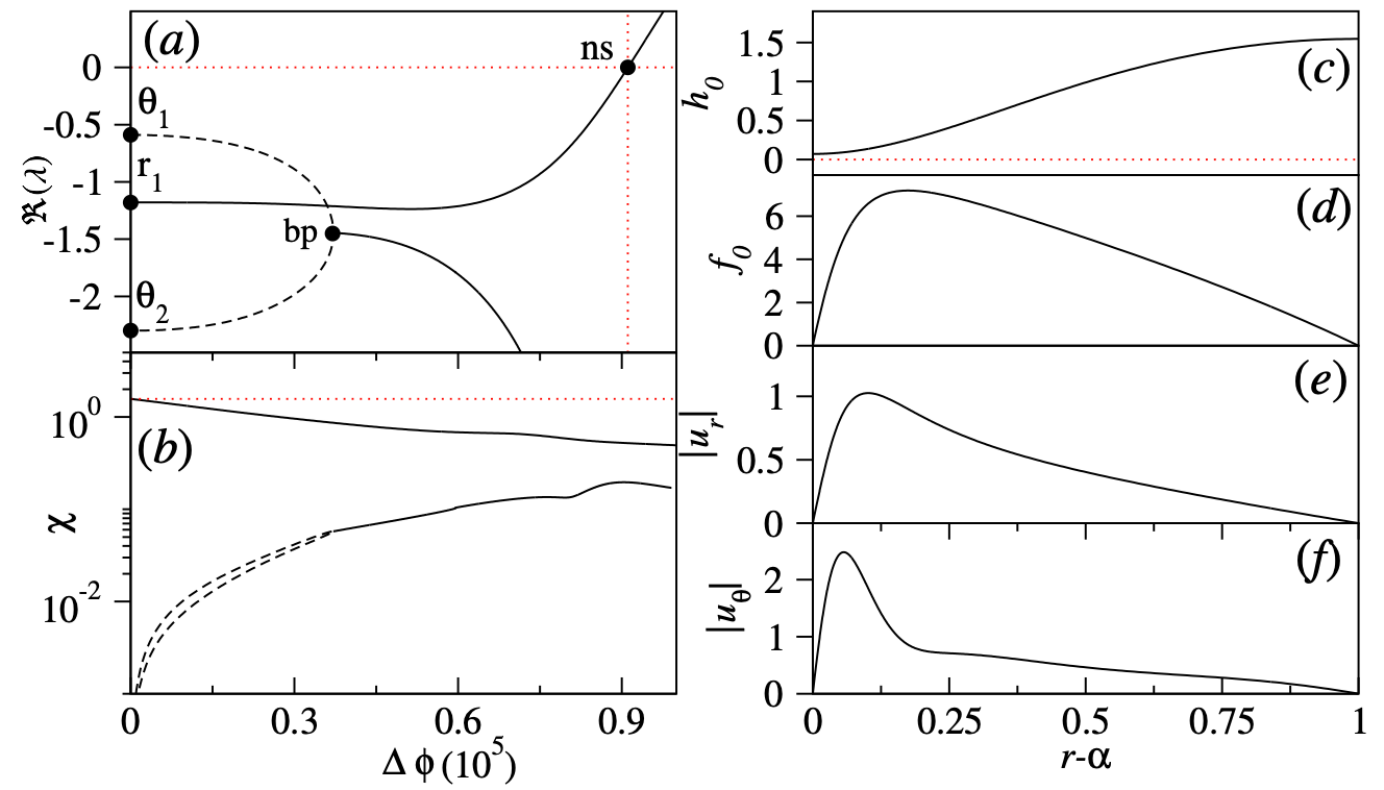}}
  \caption{($a$) Real part of the first three leading eigenvalues as a
    function of the applied voltage $\Delta\phi$ for the same
    parameters as in figure~\ref{fig2}. The solid (dashed) lines
    correspond to complex (purely real) eigenvalues,
    respectively. Labels {\tt bp} and {\tt ns} mark the loci of the
    branching point of the azimuthal velocity mode and the point of
    neutral stability of the radial velocity mode, respectively; ($b$)
    the relative strength of the radial to azimuthal velocity mode
    $\chi$; ($c$) the neutrally stable film profile at point {\tt ns} in
    panel ($a$); ($d$) the azimuthal velocity $f_0$ in a neutrally stable
    film; the magnitudes of the neutrally stable radial ($e$) and
    azimuthal ($f$) velocity perturbations.}\label{fig4}
\end{figure}
The real parts of the first three leading eigenvalues labelled by $r_1$,
$\theta_1$ and $\theta_2$ are shown in figure~\ref{fig4}($a$). At
$\Delta\phi=0$ the leading mode is the azimuthal velocity mode
$\theta_1$ with the real eigenvalue
$\lambda_{\theta_1}\approx-0.589$. The second least stable mode $r_1$
corresponds to the radial velocity. It has the complex eigenvalues
$\lambda_{r_1}=-1.179\pm10.149\ii$. The third mode $\theta_2$ is again
the azimuthal velocity mode with the real eigenvalue
$\lambda_{\theta_2}\approx-2.298$. As $\Delta\phi$ increases, the two
leading azimuthal velocity modes remain monotonically stable until the
branching point {\tt bp} is reached, where the two modes form a
complex conjugate pair with a negative real part collide. A further
increase of $\Delta\phi$ leads to the destabilisation of the leading
radial mode $r_1$, which becomes neutrally stable at the point {\tt
  ns}, where $\Im\{\lambda_{r_1}\}\not=0$. The corresponding film
thickness profile $h_0(r)$ and the base azimuthal flow $f_0(r)$ are
shown in figures~\ref{fig4}($c,d$), respectively.

Oscillatory neutrally stable perturbation is a mixture of radial
and azimuthal flows as quantified by $\chi$ in
figure~\ref{fig4}($b$). Since the perturbation fields $u_r$ and
$u_\theta$ are both complex, the real physical radial and azimuthal
perturbations of the base flow are given by the real and imaginary
parts of
$|u_r|e^{\ii(\Im\{\lambda_{r_1}\}t+\Psi_r(r)})$ and
$|u_\theta|e^{\ii(\Im\{\lambda_{r_1}\}t +\Psi_\theta(r))}$ with
spatially varying phase shifts
$\Psi_r=\arctan(\Im\{u_r\}/\Re\{u_r\})$ and
$\Psi_\theta=\arctan(\Im\{u_\theta\}/\Re\{u_\theta\})$. 
The spatially varying amplitudes of the neutrally stable radial and
azimuthal velocity perturbations $|u_r|$ and $|u_\theta|$ are shown
in figure~\ref{fig4}($e,f$).

\begin{figure} 
  \centerline{\includegraphics[width=\columnwidth]{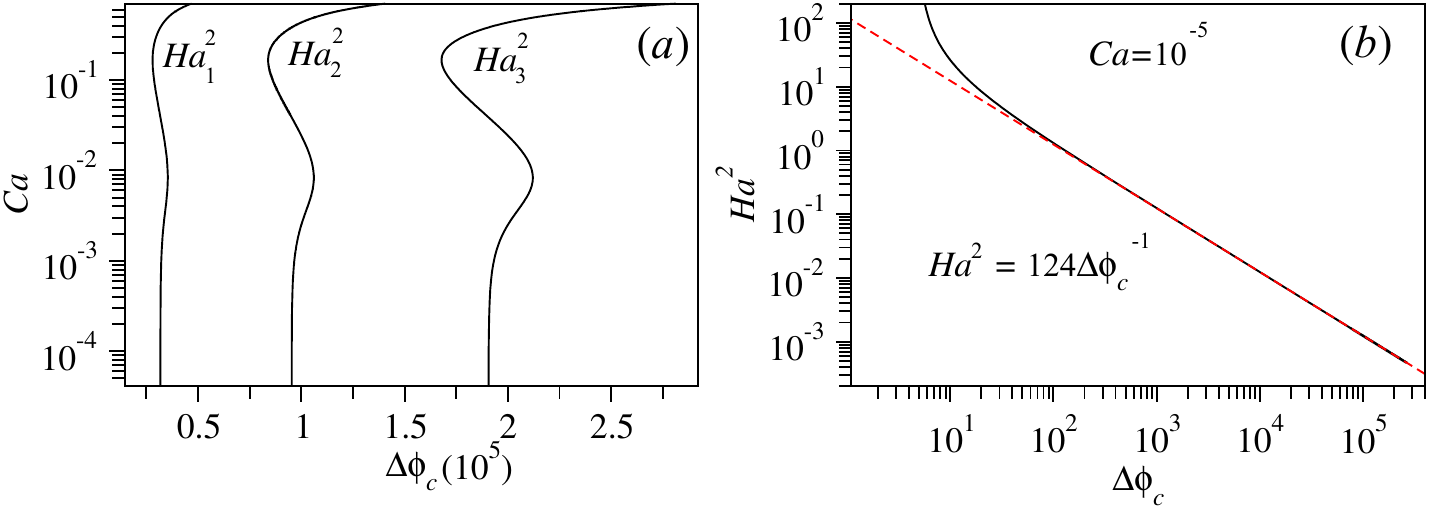}}
  \caption{($a$) Neutral stability curves in the $(\Delta\phi_c,\Ca)$
    plane for $\Ha_1^2=0.0038$, $\Ha_2^2=0.0013$ and
    $\Ha_3^2=0.00065$. ($b$) Neutral stability curve in the
    $\left(\Delta\phi_c,\Ha^2\right)$ plane for $\Ca=10^{-5}$. The
    dashed line corresponds to the power law function
    $\Ha^2=124\Delta\phi_c^{-1}$.\label{fig5}}
\end{figure}

To gain a deeper understanding of how the stability of the base flow
changes with other parameters we plot neutral stability curves in
figure~\ref{fig5}($a$) for three selected values of $\Ha^2$ in the
plane $(\Delta\phi_c,\Ca)$, where $\Delta \phi_c$ is the critical
value of the applied voltage above which the base flow is linearly
unstable. It is noteworthy that for each value of $\Ha^2$ there exists
a finite limiting value of $\Delta\phi_c$ as $\Ca\to0$. Such a
limiting behaviour of the neutral stability curves confirms the
dynamic nature of the flow instability. Indeed, with the scaling used 
here the capillary number is given by $\Ca=\mu(\gamma\langle
h\rangle\rho)^{-1/2}$ and the physical value of the applied voltage is
\[V_c=UB(R_2-R_1)\Delta\phi_c
  =B\sqrt{\frac{\gamma\langle h\rangle}{\rho}}\Delta\phi_c\,.\]
For the fixed $\langle h\rangle$, viscosity $\mu$ and density $\rho$
the limit of $\Ca\rightarrow0$ corresponds to a large surface tension
$\gamma\to\infty$. The existence of a finite value of $\Delta \phi_c$
implies that regardless of the strength of stabilising surface tension
forces the instability can always be induced if the applied voltage
exceeds $V_c$.

We set $\Ca=10^{-5}$ and trace the locus of the neutral stability
in the $(\Delta\phi_c,\Ha^2)$ plane in figure~\ref{fig5}($b$). The
dashed line corresponds to $\Ha^2=124\Delta\phi_c^{-1}$ and is
obtained by fitting the lower part of the curve (small $\Ha^2$) with a
power law function. This simple power law relationship between $\Ha^2$
and $\Delta\phi_c$ represents a universal stability threshold in the
limit of small capillary and Hartmann numbers. Reintroducing
dimensional variables we obtain the critical value of the applied
voltage $V_c$ 
\begin{equation}
  V_c=\frac{124\mu}{B\sigma (R_2-R_1)^2}
  \sqrt{\frac{\gamma\langle h\rangle}{\rho}}\,.\label{thresh} 
\end{equation}
The asymptotic result \ref{thresh} is valid for the selected radii
ratio $R_2/R_1=2$ and $\left(\Ca,\Ha^2\right)\to(0,0)$. At the critical
value of $V_c$ the film is strongly deformed and has a shape similar
to the solution (2) in figure~\ref{fig2}($c$). The minimum value of
the film thickness is attained at the inner cylinder and is
approximately 6\% of $\langle h\rangle$.

\section{Conclusions} \label{concl}
We derived a set of leading order equations in the lubrication
approximation that describe the electromagnetically driven flow of
electrolyte solutions and/or non-magnetic liquid metals in a thin
horizontal free layer with deformable surfaces placed in a uniform
magnetic field normal to the layer. The equations are written in a
generic coordinate-invariant form and can be used to study
electromagnetically driven flows in an arbitrary geometry dictated by
the shape and size of the deployed electrodes. The equations account
for the presence of surfactants and chemical species dissolved in the
bulk of the fluid. Inspired by recent studies of the
electromagnetically driven flows in supported shallow annular layers
between two coaxial cylinders \cite{Figueroa09, Perez16, Suslov17,
  McCloughan20}, we choose a similar geometry and apply the derived
model to investigate the flow and its stability in annular free film
spanning the gap between two coaxial cylindrical electrodes. 

Similar to mechanically driven flows in soap films \citep{Wu95},  a
steady azimuthal electromagnetically driven flow can only exist 
in a deformed layer, where the radial component of the gradient of the
Laplace pressure balances the centrifugal force. This is in contrast
to supported thicker layers, where the free-surface deformation is
typically negligible compared to the thickness of the layer while the
flow speed does not exceed several centimetres per second. The other
important feature distinguishing the flows in free layers and films
from those in supported layers is that in the former the Laplace
pressure dominates while in the latter the hydrostatic pressure
gradient defines the fluid trajectory in the plane of the layer.

If the intermolecular forces are neglected, the azimuthally invariant  
steady-state flow can be found for an arbitrarily large electric
current flowing through the film. The minimal film thickness is
achieved at the inner electrode. It decreases as a power law function
of the applied voltage remaining non-zero so that the point of a true
film rupture is never reached. 
 
We determined that the azimuthal flow in approximately flat free films
with a small velocity is linearly unconditionally stable. For
relatively weak magnetic fields of the order of $10^{-2}$\,T and fluid
parameters corresponding to a weak electrolyte solution used in the
experiments of \cite{Figueroa09, Perez16}, we found that the azimuthal
velocity perturbations decay monotonically while the decay of radial
velocity perturbations of the base flow in layers with the average
thickness in the micrometre range is oscillatory. However, as the film
thickness is decreased below a certain critical value given by
condition~\ref{ls_osc} ($\lesssim100$\,nm for a film existing
between coaxial cylinders with the inner and outer radii of 1 and
2\,cm, respectively), the relaxation dynamics is found to be dominated
by viscous damping with a monotonic decay of a radial flow and the
associated surface deformation. This result contrasts the observations
of instabilities in supported thicker annular layers, where the
primary flow becomes unstable with respect to three-dimensional
perturbations of the velocity without a noticeable variation of the
layer depth. The fluctuations of the solute concentration in the bulk
has no effect on the growth rate of the leading modes. In the presence
of a surfactant, the Marangoni effect leads to further stabilisation
of the base flow, which is in agreement with earlier studies of planar
soap films in the absence of magnetic fields \citep{Miksis98}.

By following the branch of steady-state solutions into the regime of
large applied voltage and, consequently, large electric currents, we
find that the steady azimuthal flow eventually becomes unstable with
respect to a mixture of oscillatory azimuthal and radial velocity
perturbations at a certain critical value of the applied voltage, at
which the deformation amplitude of the layer is of the order of the
average film thickness. This suggests that electromagnetically driven
flows in free films may not be ideal candidates for studying
two-dimensional turbulence since the primary instability of the base
flow sets in only when the film is already strongly deformed. We note
that a similar conclusion was made earlier in \cite{Rivera00}, where
it was hypothesised that strong damping caused by the interactions of
the flow within a film with a surrounding gas layer that are enhanced
by the surface deformation may be the cause for the energy
leakage responsible for stronger than expected decays of the velocity
correlation function.

Finally, we briefly mention the action of intermolecular forces
characterised by the disjoining pressure. They play a major role in
the instability and subsequent stabilisation of free soap films that
are thinner than $\sim100$\,nm. It is well known
\cite{Israelachvili11, Overbeek60} that long-range van der Waals 
forces destabilise free layers sandwiched between dielectric media
with identical properties. These forces alongside the gravity-induced
drainage of the fluid constitute the primary source of film
instability. As the film thickness decreases below$10-50$\,nm, an
electric double layer is typically formed consisting of the monolayers
of soap ions adsorbed at each interface. Strong electric double-layer
forces lead to the repulsion between the film surfaces and the
formation of highly stable black soap films with a thickness under
$50$\,nm. Therefore, it is important to study the role of the
disjoining pressure on the steady azimuthal flow and its stability in
the strong-current regimes when the deformation of the layer is
significant. Intermolecular forces, black soap films and film rupture
will become the topic of our future investigations.

\bibliographystyle{jfm}
%\bibliography{MHDFilmRef}

\end{document}